\documentclass[%
aps,pra,twocolumn,superscriptaddress
]{revtex4-1}
\usepackage{graphicx}
\usepackage{subfigure}
\usepackage{dcolumn}
\usepackage{bm}
\usepackage{hyperref}
\usepackage{indentfirst}
\usepackage{braket}
\usepackage{amsmath}
\usepackage{xcolor}
\hypersetup{colorlinks=true, citecolor=blue, urlcolor=blue, linkcolor=blue}
\pdfoutput=1

\begin{document}
\title{Condensation of Composite Bosonic Trions in Interacting Bose-Fermi Mixtures}
\author{Qi Song}
\author{Jie Lou}
\email{loujie@fudan.edu.cn}
\author{Yan Chen}
\email{yanchen99@fudan.edu.cn}
\affiliation{Department of Physics and State Key Laboratory of Surface Physics, Fudan University, Shanghai 200433, China}

\date{\today}

\begin{abstract}
We reveal a quantum coherent state characterized by composite bosonic trions, wherein paired fermions further bind with bosons, in one-dimensional Bose-Fermi mixtures.
This phase emerges in two separate models, both featuring onsite boson–fermion attraction that induces negative binding energy for the composite trions. 
The first is the pair–hopping model, in which spinless fermions undergo pair hopping to form preformed pairs. The second is the extended Bose-Fermi Hubbard model, describing dipolar bosons and fermions with density–density interactions. 
Notably, the formation of composite trions is independent of the specific pairing interactions of fermions.
Extensive density matrix renormalization group calculations demonstrate the quasi-condensation of the composite trions, evidenced by algebraically decaying correlations of trions, gapped single-particle excitations, and suppressed fermion pair correlations. 
Our findings provide valuable insights into the three-body pairing mechanism in mixed-particle systems.

\end{abstract}
\maketitle

\section{Introduction}

Three-body bound states, often referred to as trimers or trions, are fundamental entities that span diverse fields such as nuclear and particle physics, condensed matter physics, and quantum optics \cite{hammer2010efimov,RevModPhys.85.1633,Naidon_2017}. 
For instance, three bound quarks form neutrons and protons \cite{peskin2018introduction}.
In few-body physics, bosonic trimers associated with the well-known Efimov effect have been extensively studied \cite{Efimov1973,4HeTrimerFewBody2017,4He3Efimov2015}.
Fermionic trimers have also attracted significant interest in interacting ultracold Fermi gases \cite{fermionicTrimer2009exp,fermionicTrimer2011exp,fermionicTrimer2014HuiZhai,ferimionicTrimer2010}.
In addition, solid-state Bose-Fermi mixtures, such as two-dimensional transition metal dichalcogenides, have emerged as promising platforms for exploring fermionic trions, where excitons (bound electron-hole pairs) bind with free electrons or holes via electrically tunable Feshbach resonances \cite{TMDexp2021,TMDtheory2025}.
Remarkably, at room temperature, Bose-Einstein condensates of exciton-polaritons have been realized in semiconductor and organic microcavities \cite{Yoshihisa2014polariton,keeling2020polariton,ghosh2022polariton}. Polaritons are hybrid bosonic quasiparticles representing a superposition of an exciton (two fermion) and a cavity photon (one boson). Due to their finite lifetime, polaritons inherently possess a non-equilibrium nature. This naturally prompts a fundamental inquiry: in equilibrium, is there a comprehensive theoretical framework elucidating how a fermion pair and a boson can bind to form a composite bosonic trion, yielding a coherent and stable ground state? Furthermore, can such an exotic phase of matter be realized through quantum simulation?

Over the past two decades, ultracold bosonic and fermionic atoms have proven to be exceptionally versatile platforms for simulating quantum physics \cite{RevModPhys.80.885,doi:10.1126/science.aal3837,Tools}. 
Since atomic mixtures are highly controllable, rich phenomena have also been observed in Bose-Fermi mixtures (BFM), including polarons \cite{atoms10020055polaron,baroni2024polaron,BFMpolaron2021exp}, 
heteronuclear molecules \cite{PhysRevLett.93.143001,milczewski2023molecules,duda2023molecules}, and double superfluids \cite{doi:10.1126/science.1255380, PhysRevLett.118.055301,doubleSF2016}. Beyond extensive investigations of composite fermions (a fermion plus a boson) \cite{PhysRevLett.93.143001,milczewski2023molecules,duda2023molecules,bfmPhaseDiagram2006,PhysRevLett.91.150403,PhysRevLett.93.120404,PhysRevA.77.012115,PhysRevA.77.023601,PhysRevA.77.023608}, previous theoretical work has analyzed the formation of fermionic trios, comprising a composite fermion and an elementary boson, under three-particle interaction using a diagrammatic formalism \cite{trio2004}. 
Recently, an ultracold gas of $^{23}Na^{40}K_2$ triatomic molecules has been successfully created through an atom–diatomic-molecule Feshbach resonance between fermionic $^{23}Na^{40}K$ molecules and fermionic $^{40}K$ atoms \cite{Pan2022CreationAtomMoleculeNaKK,Pan2022EvidenceAtomMoleculeNaKK,Pan2024PhotoassociationNaKK}.
These advances in quantum simulation underscore the potential for realizing an exotic pairing process, wherein fermion pairs and bosons bind to form novel composite bosonic trions that can condense. Nevertheless, a comprehensive investigation into the underlying pairing mechanism for composite bosonic trions is still lacking.

In this paper, we investigate the formation of composite bosonic trions, wherein a fermion pair further associates with a boson.
We first explore the composite bosonic trion phase within the Bose–Fermi model incorporating fermion pair hopping, where the kinetic pair tunneling of spinless fermions \cite{cc32spinless2017,TwoFluid2021,KineticFpair2022} induces preformed fermion pairs.
Numerical evidence reveals that, under onsite boson–fermion attraction, only the trion correlation function displays algebraic decay, indicative of a single gapless bosonic mode with a central charge $c=1$. The quasi-condensation momentum of the composite trions is dictated by the sign of the fermionic pair-hopping amplitude. Furthermore, the stability of the composite trions is reinforced by negative binding energy, gapped single-particle excitations, and suppressed correlations among paired fermions. Our calculations also show that, as the boson–fermion interaction weakens, the composite trion phase transitions smoothly into a two-component Luttinger liquid (TLL), where fermion pairs and bosons each contribute a gapless excitation mode, resulting in an overall central charge of $c=2$. 

Building on insights gained from the preceding pair–hopping model, we generalize our analysis to the extended Bose-Fermi Hubbard model 
\cite{Zoller2012dipolarReview,EBFH2010Wang,PhysRevA.81.053626,EBFH2025SciPost}, which describes dipolar bosons and fermions interacting via both short-range onsite attractions and long-range density–density interactions (including first- and second–neighbor couplings). Within this framework, we identify a similar composite trion phase that occupies a substantial portion of the phase diagram. 
Remarkably, the composite trion formation does not depend on the particular physical origin underlying fermion pairing, emphasizing the universal nature of the trion formation mechanism.
Our work establishes a versatile platform for exploring intriguing three-body bound states, which complement the well-known bosonic trimers in Bose systems and fermionic trimers in Fermi systems, and hold promise for diverse applications in quantum many-body physics and quantum chemistry \cite{carr2009coldApp,bohn2017Chemistry}.

\section{Composite bosonic trions in the pair–hopping model}
The first model we consider incorporates the typical onsite boson-fermion interaction of strength $U_{bf}$ \cite{bfmPhaseDiagram2006,PhysRevLett.91.150403,PhysRevLett.93.120404,PhysRevA.77.012115,PhysRevA.77.023601,PhysRevA.77.023608} and a kinetic component for fermions that exhibits competition between single and paired configurations, as introduced in Ref. \cite{TwoFluid2021,pairHopping2021spinless,pairHopping2022spinless,bariev1991spinless}.
The Hamiltonian in a chain of size $L$ is given by
\begin{equation}\label{Ham}
\begin{split}
H=& -\sum_{i} (t_fc^{\dagger}_ic_{i+1} +t_bb^{\dagger}_ib_{i+1}+H.c.) + U_{bf}\sum_{i}n^b_{i}n^f_{i}  \\
 &-t_{ff}\sum_{i} (c^{\dagger}_{i+1}c^{\dagger}_ic_ic_{i-1}+H.c.)
+\frac{U_{bb}}{2}\sum_{i} n^b_{i}(n^b_{i}-1),
\end{split}
\end{equation} 
where $c^{\dagger}_i$ ($b^{\dagger}_i$) creates a spinless fermion (a boson) at site $i$, $t$ and $t_{ff}$ denote the tunneling rates for single particles and paired fermions respectively, and $n^b_{i}=b^{\dagger}_ib_i (n^f_{i}=c^{\dagger}_ic_i)$ represents the boson (fermion) number operator. 
$U_{bb}$ denotes the onsite repulsive interaction between bosons. We set $t_f=t_b=t=1$ as the unit of energy and $t_{ff}=3.0$ unless otherwise specified. The fermion density is fixed at $\rho_f=N_f/L=1/4$, with a fermi-boson ratio $N_f/N_b$ set to $2\colon 1$. 
In our density matrix renormalization group (DMRG) \cite{PhysRevLett.69.2863,ITensor}calculations, we consider system sizes up to $L=160$ with open boundary conditions (OBC) and $L=96$ with periodic boundary conditions(PBC). The maximum boson occupation number per site, $n^{cut}_b$, is set to 2 due to the boson filling $\rho_b = N_b/L = 1/8$ and the large onsite repulsion $U_{bb} = 8.0$. Calculations indicate that further increasing $n^{cut}_b$ does not alter our results.

\begin{figure}
\centering
\includegraphics[scale=0.74,trim=10 15 1 0,clip]{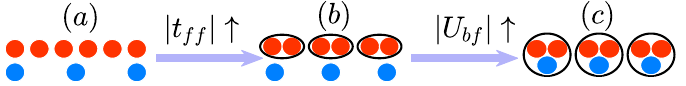}
\caption{\label{fig_Schematics}Schematics for the formation of composite trions in a mixture of fermions (red) and bosons (blue) with a number ratio of $2\colon 1$. As the tunneling rate for paired fermions $t_{ff}$ increases, individual fermions neighboring each other transition from a single state (a) to a paired state (b). By further increasing the strength of the attractive interaction $|U_{bf}|$ between fermions and bosons, the system evolves continuously from a two-component Luttinger liquid consisting of fermion pairs and bosons (b) to a single-component Luttinger liquid composed of composite trions (c).
}
\end{figure}
\subsection{Composite bosonic trion formation}
At large values of $|t_{ff}/t_f|$, it has been shown \cite{TwoFluid2021} that the fermions fully pair up. In such a scenario, the system exhibits only one gapless bosonic excitation mode with a central charge $c=1$. These paired fermions quasi-condensate around either $k=0$ or $k=\pi$ momenta, depending on the sign of $t_{ff}$. When bosons are introduced into the system and two-body contact interactions $U_{bf}$ are considered, the paired fermions are attracted to the bosons, leading to the formation of composite trions, as illustrated in Fig. \ref{fig_Schematics}. For sufficiently strong attractive interactions, $U_{bf} \ll -|t_{ff}|$, the system can be described by an effective model derived from perturbation theory \cite{perturbation2003,perturbationUpSSH2021}. The effective Hamiltonian is given by

\begin{eqnarray} \label{Heff}
H_{eff}=-\frac{2t_{ff}t_b}{|U_{bf}|}\sum_{i} ( c^{\dagger}_{i+1}b^{\dagger}_{i+1}c^{\dagger}_ic_ib_ic_{i-1}+H.c).
\end{eqnarray}
Therefore, in this limit, it is anticipated that the system is predominantly composed of composite trions.

\begin{figure}
\centering
\includegraphics[scale=0.75,trim=2 0 0 0,clip]{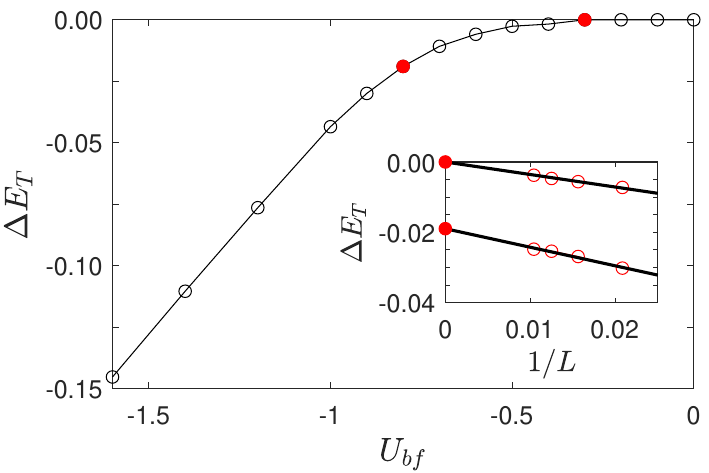}
\caption{\label{fig_DE_Ubf}Binding energy of trions versus $U_{bf}$ for $t_{ff}=3.0$. The black circles depict the extrapolated results approaching the thermodynamic limit as $1/L\to 0$. The inset shows the finite-size values (open red circles) of the binding energy obtained from systems with sizes $L=48,64,80,96$ with PBC, and their extrapolation (solid red circles, both in the inset and the main panel) by a linear fitting at typical $U_{bf}=-0.3,-0.8$.  
}
\end{figure}

An important energy criterion for identifying the formation of composite trions is the trion binding energy. It is defined as follows, 
\begin{eqnarray} \label{eq_DE}
\Delta E_T=\lim\limits_{L\to\infty} [ E_L(N_f+2, N_b+1)+E_L(N_f,N_b)\nonumber \\ 
 -E_L(N_f+2, N_b)-E_L(N_f,N_b+1)],
\end{eqnarray}
where $E_L(N_f,N_b)$ represents the ground-state energy of a system with $N_f$ fermions and $N_b$ bosons in a chain of length $L$. In the scaling analysis, we fix $\rho_f=2\rho_b=1/4$. The binding energy $\Delta E_T$ measures the energy difference between a composite trion bound state and its two unbound components. In the thermodynamic limit, a negative $\Delta E_T$ indicates a strong tendency for the formation of composite trions. Fig. \ref{fig_DE_Ubf} shows that $\Delta E_T$ begins to drop below zero around $U_{bf}=-0.4$, even before the system enters the composite trion phase. This suggests that composite trions begin to form before phase coherence is established. In the deep composite trion phase ($U_{bf}<-1.0$), the magnitude of $\Delta E_T$ increases proportionally with the attraction strength $U_{bf}$, indicating stronger binding energy for the trions as the interaction becomes more attractive.

\begin{figure}
\centering
\includegraphics[scale=0.62,trim=0 0 0 0,clip]{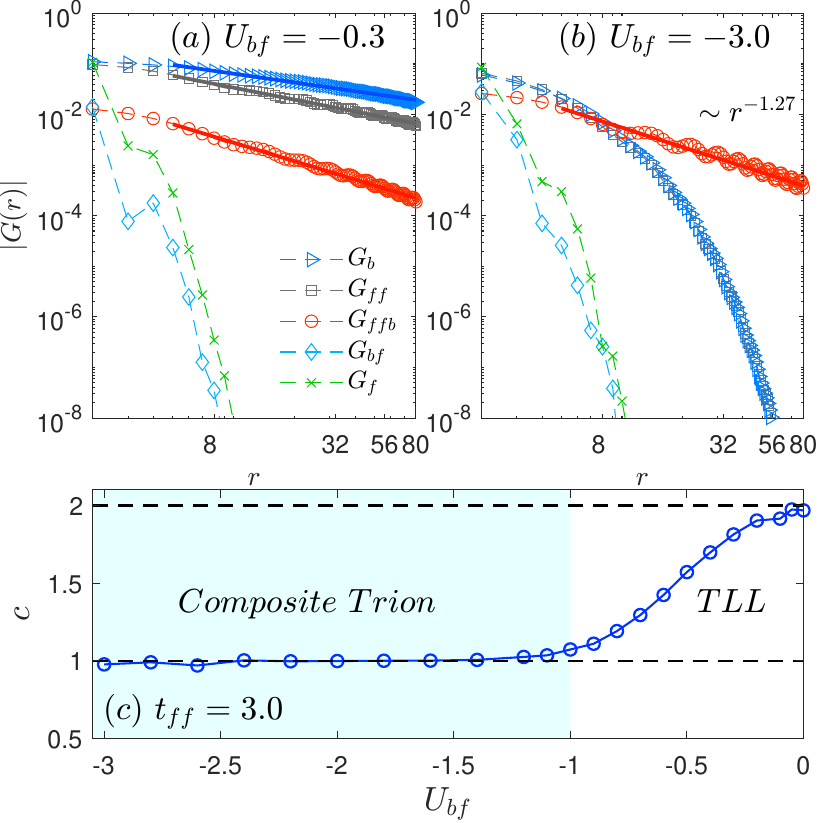}
\caption{\label{fig1_Gr_cc} Correlation functions for (a) TLL phase at $U_{bf}=-0.3$ and (b) composite trion phase at $U_{bf}=-3.0$ in a 1D system of length $L=160$ with OBC. $t_{ff}=3.0,\rho_f=2\rho_b=1/4$ are fixed and $r$ is the distance from the reference site at $L/4$. A double-logarithmic scale is used in both (a) and (b) for clarity. The solid lines denote power-law fittings to $|G(r)|\sim r^{-\alpha_q}$, yielding exponents of $\alpha_b= 0.57, \alpha_{ff}= 0.78, \alpha_{ffb}= 1.21$ in (a) and $\alpha_{ffb}= 1.27$ in (b).  
(c) Central charge $c$ as a function of $U_{bf}$, obtained by fitting the von Neumann entropy. Within the composite trion phase, $c=1$ when $U_{bf}\leq -1.0$.
}
\end{figure}

\subsection{Correlation functions and momentum distributions}To fully characterize the quantum phases and determine the quasi-long-range order (QLRO) within our system, we evaluate the following correlation functions:
\begin{equation}\label{eq_Gr}
    \begin{split}
&G_b(r)= \langle b^{\dagger}_ib_{j} \rangle,  G_f(r)= \langle c^{\dagger}_ic_{j} \rangle, \\
&G_{bf}(r)= \langle c^{\dagger}_ib^{\dagger}_ic_{j} b_{j} \rangle,  
G_{ff}(r)= \langle c^{\dagger}_ic^{\dagger}_{i+1}c_{j}c_{j+1} \rangle,\\
&G_{ffb}(r)= \langle c^{\dagger}_ib^{\dagger}_ic^{\dagger}_{i+1}c_{j}b_{j}c_{j+1} \rangle,
    \end{split}
\end{equation}
where $r=|i-j|$ is the distance separating sites $i$ and $j$.
The QLRO is determined by the power-law decay exhibited by the dominant correlation function.
Notably, genuine long-range order is absent in 1D systems in the thermodynamic limit. These correlation functions take the simplest forms, either power-law decay $\sim r^{-\alpha_q} $ or exponential decay $\sim e^{-r/d_q} $, 
with $q$ denoting the particle species (boson $b$, fermion $f$, paired fermions $ff$, composite fermion $bf$ and composite bosonic trion $ffb$). In the context of  Luttinger liquid theory, the exponents are given by $\alpha_{b(ff)}=1/(2K_{b(ff)})$,  and $ \alpha_{f(bf)}=(K_{f(bf)}+1/K_{f(bf)})/2$, where $K_q$ are the Luttinger parameters  \cite{PhysRevA.77.023621,bfmPhaseDiagram2006,fpaired2019}.
At zero temperature, the predominant quantum phase is determined by the correlation function with the slowest decay, characterized by the minimal positive power exponent $\alpha_{q}<2$, which signifies a divergent susceptibility \cite{Kivelson2004,Mathey2007polaron}.

\begin{figure}
\centering
\includegraphics[scale=0.73,trim=4 0 1 0,clip]{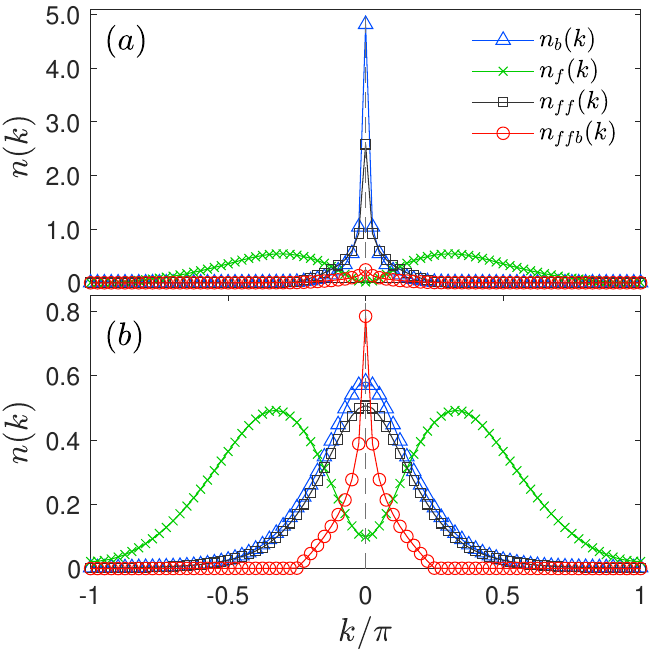}
\caption{\label{fig_nk_tff-}Momentum distribution functions $n(k)$ for  $t_{ff}=-3.0$ for (a) the TLL phase at $U_{bf}=-0.3$ and (b) the composite trion phase at $U_{bf}=-6.0$ in the $L=80$ system with PBC. The grey dashed lines guide the eyes at $k=0$.
}
\end{figure}
When the onsite attraction between fermions and bosons is relatively weak, the paired fermions exhibit only a weak coupling to the bosons. In this scenario, the system behaves as a two-component Luttinger liquid, comprising fermion pairs and bosons. Within this framework, the correlations of bosons, paired fermions, and composite trions all decay according to a power law $\sim r^{-\alpha}$, while the single-particle correlations of fermions and composite fermions decay exponentially, as shown in Fig. \ref{fig1_Gr_cc}(a). The correlations of the trions can be understood as a direct combined projection of the correlations of the two components, arising due to the presence of $U_{bf}$ interaction. 
Consequently, this TLL phase exhibits only two gapless bosonic modes. In accordance with (1+1)-dimensional conformal field theory (CFT) \cite{fpaired2019,ShoushuGong2023,Calabrese_2009,Calabrese_2004}, this implies that the central charge should be 2. Indeed, our analysis confirms that the central charge $c\approx 2$, as determined by the von Neumann entanglement entropy at small $|U_{bf}|$ (see \ref{AppendixB}).

Upon further increasing the magnitude of the two-body attraction $|U_{bf}|$, the correlation of composite trions emerges as the sole correlation decaying according to a power law, which can be observed in Fig. \ref{fig1_Gr_cc}(b). In stark contrast to the TLL phase, the correlation functions of paired fermions $G_{ff}$ undergo exponential suppression. Notably, single-particle excitations exhibit gaps, evidenced by the exponential decay of both single-fermion ($G_{f}$) and single-boson ($G_{b}$) correlations. 
Given the central charge $c=1$, it becomes evident that the elementary particle within the system is the composite bosonic trion, consisting of two fermions and one boson, which contributes an effective bosonic mode.  
As $U_{bf}$ varies, the central charge evolves continuously, as depicted in Fig. \ref{fig1_Gr_cc}(c). When the system undergoes a complete transformation into the composite trion phase, the central charge attains a value of 1 for $U_{bf}\leq -1.0$.  

The Fourier transforms of the correlation functions yield the momentum distributions $n_q(k)= \frac{1}{L}\sum_{i,j}G_q(i,j)e^{ik(i-j)}$ for different particles $q$. In the TLL phase, paired fermions exhibit quasi-condensation around momentum $k=0$ at $t_{ff}=-3.0$ (alternatively, around $k=\pi$ at $t_{ff}=+3.0$, as shown in \ref{AppendixB}). 
Similarly, the momentum distribution of bosons displays a sharp peak $k=0$, as depicted in Fig. \ref{fig_nk_tff-}(a). On the contrary, in the composite trion phase, the momentum distributions of both paired fermions and bosons are significantly broadened. 
Notably, only the composite trions exhibit a prominent peak in momentum space, which is visible in Figure \ref{fig_nk_tff-}(b). The condensation momentum of composite trions is $\pi$ for positive $t_{ff}$ and $0$ for negative $t_{ff}$, depending on the primary distribution of the constituent fermion pairs.

\begin{figure}
\centering
\includegraphics[scale=0.63,trim=2 1 1 0,clip]{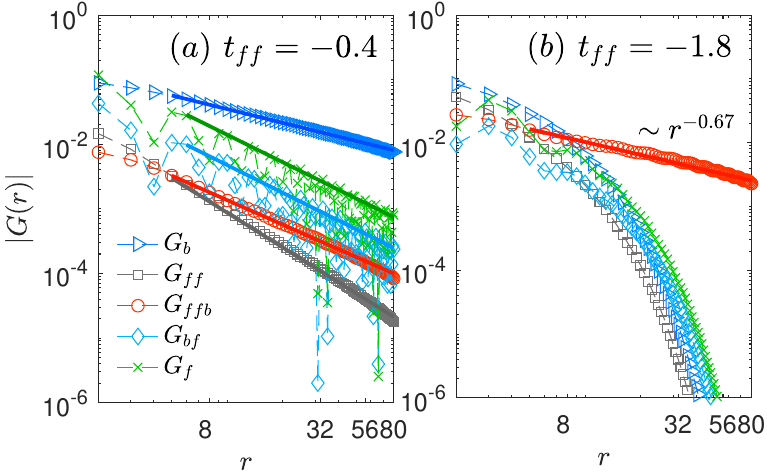}
\caption{\label{figS_Gr_tff_Ubf-3} Correlation functions with $U_{bf}=-3.0$ fixed for (a) the TLL phase at $t_{ff}=-0.4$ and (b) the composite trion phase at $t_{ff}=-1.8$ in a 1D system of length $L=160$ with OBC. $\rho_f=2\rho_b=1/4$ are fixed, and $r$ represents the distance from the reference site at $L/4$.  A double-logarithmic scale is used in both (a) and (b). The solid lines denote power-law fitting to $|G(r)|\sim r^{-\alpha_q}$ where $\alpha_b\simeq 0.70,\alpha_f\simeq 1.41,\alpha_{bf}\simeq 1.42, \alpha_{ff}\simeq 1.78, \alpha_{ffb}\simeq 1.26$ in (a) and $\alpha_{ffb}\simeq 0.67$ in (b).  
}
\end{figure}

\subsection{Influence of magnitude of $t_{ff}$} 
At small $|t_{ff}/t_f|$, another type of two-component Luttinger liquid phase involving bosons and unpaired fermions, referred to as TLLBF, emerges. 
In this phase, the two gapless modes arising from the bosonic and unpaired fermionic components yield a central charge $c=2$. 
All relevant correlation functions—$G_f,G_b,G_{bf},G_{ff}$ and $G_{ffb}$—exhibit power-law decay (see Fig. \ref{figS_Gr_tff_Ubf-3}). 
This behavior contrasts with that in the TLL phase at small $|U_{bf}|$ and large $|t_{ff}|$, where $G_f$ and $G_{bf}$ decay exponentially, while $G_{ff}$, $G_b$, and $G_{ffb}$ exhibit algebraic decay. A qualitative sketch of the ground-state phase diagram as a function of $t_{ff}$ and $U_{bf}$ is provided in \ref{AppendixB}.

When $|t_{ff}/t_f|>1.6$ and the attractive $U_{bf}$ is not negligible, the composite trion phase emerges, with only composite bosonic trions displaying QLRO. 
This indicates the importance of preformed fermion pairs in the formation of composite bosonic trions, despite the absence of QLRO for paired fermions within the composite trion phase.

\begin{figure}
\centering
\includegraphics[scale=0.55,trim=0 0 0 0,clip]{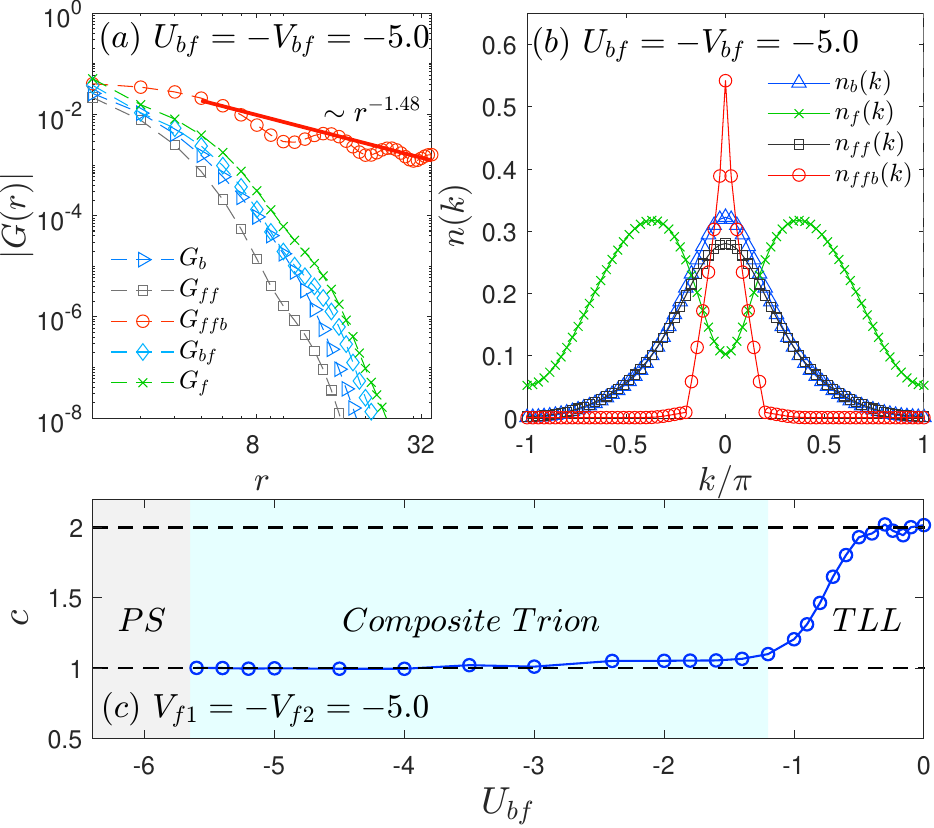}
\caption{\label{figVnnnGrCCnk} Results for the extended Bose-Fermi Hubbard Model (\ref{HamVffVbf}) in an $L=70$ system with PBC. (a) Correlation functions in the composite trion phase at $U_{bf}=-V_{bf}=-5.0$, plotted on a double-logarithmic scale. The solid line represents a power-law fit of the form $|G_{ffb}(r)|\sim r^{-\alpha_{1.48}}$. (b) Momentum distribution functions $n(k)$ at $U_{bf}=-V_{bf}=-5.0$. Only the composite trions exhibit a prominent peak at zero momentum.  
(c) Central charge $c$ as a function of $U_{bf}$. Within the composite trion phase, $c=1$ when $-5.8\leq U_{bf}\leq -1.2$. "PS" denotes phase separation, characterized by randomly distributed particle clusters lacking a well-defined central charge.
For all panels, $V_{f1}=-V_{f2}=-5.0,V_{bf}=-U_{bf},U_{bb}=8.0,t_f=t_b=1$, and $\rho_f=2\rho_b=1/5$ are fixed.
}
\end{figure}
\section{Composite bosonic trions in the Extended Bose-Fermi Hubbard Model}
At this point, our analysis based on the pair–hopping model has underscored the pivotal role of preformed fermion pairs in driving the composite bosonic trion phase. Motivated by these insights, we now extend our investigation to a more general density-density interaction framework. 
In particular, we investigate the extended Bose–Fermi Hubbard model \cite{Zoller2012dipolarReview,EBFH2010Wang,PhysRevA.81.053626,EBFH2025SciPost}, which describes a BFM of dipolar particles interacting via both short-range and long-range dipole–dipole interactions.
This model incorporates both first-neighbor ($V_{f1}$) and second-neighbor ($V_{f2}$) density-density interactions between fermions. Moreover, bosons and fermions interact via onsite ($U_{bf}$) and nearest-neighbor ($V_{bf}$) density-density interactions. 
The extended Bose–Fermi Hubbard Hamiltonian reads
\begin{eqnarray}\label{HamVffVbf}
H_{E}=\!\!\!\!\!\!&& -\sum_{i} (t_fc^{\dagger}_ic_{i+1} +t_bb^{\dagger}_ib_{i+1}+H.c.)  + U_{bf}\sum_{i}n^b_{i}n^f_{i} \nonumber \\
 \!\!\!\!\!\!&& + \frac{V_{bf}}{2}\sum_{i}(n^b_{i}n^f_{i+1} +n^b_{i+1}n^f_{i} )+\frac{U_{bb}}{2}\sum_{i} n^b_{i}(n^b_{i}-1) \nonumber \\
\!\!\!\!\!\!&& + V_{f1}\sum_{i} n^f_{i}n^f_{i+1} + V_{f2}\sum_{i} n^f_{i}n^f_{i+2}.
\end{eqnarray}
This extended model not only captures the essential physics revealed by the analysis based on the pair–hopping model (\ref{Ham}), but also provides a versatile platform for exploring the stability and robustness of the composite bosonic trion phase. 

\begin{figure}
\centering
\includegraphics[scale=0.7,trim=0 0 0 0,clip]{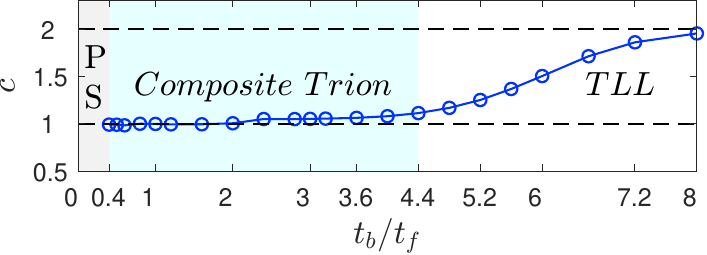}
\caption{\label{figccVfftb}Central charge $c$ as a function of $t_b/t_f$ in an $L=70$ system with PBC. Within the composite trion phase, $c=1$ for $0.4\leq t_b/t_f\leq 4.4$. "PS" denotes phase separation for $t_b/t_f<0.4$, characterized by randomly distributed particle clusters lacking a well-defined central charge. The parameters are set as $V_{f1}=-V_{f2}=-5.0,U_{bf}=-V_{bf}=-5.0,U_{bb}=8.0$, and $\rho_f=2\rho_b=1/5,t_f=1$.
}
\end{figure}

The fermionic sector of the Hamiltonian (\ref{HamVffVbf}) is a paradigmatic setup in one-dimensional lattice models of spinless fermions  \cite{spinlessFDD32,spinlessFDD33,spinlessFDD34,spinlessFDD41,fpaired2019,fpairingPhysRevResearch2021}. Its ground state supports a cluster Luttinger liquid of paired fermions with gapped single-fermion excitations when $V_{f1}<0$ and $V_{f2}>0$ at fermion densities below half-filling. We set $t_f=1,V_{f1}=-V_{f2}=-5.0$ and $\rho_f=1/5$, a typical parameter choice in the pair cluster Luttinger liquid, to facilitate the formation of preformed fermion pairs. The attraction between bosons and fermions can drive the system into the composite trion phase.
As shown in Fig. \ref{figVnnnGrCCnk}, the trion correlation function $G_{ffb}$ exhibits QLRO with power-law decay, while the trion occupation number $n_{ffb}(k)$ features a pronounced peak at zero momentum. As the boson-fermion attraction strengthens, the system transitions from the TLL phase to the composite trion phase, as reflected in the central charge behavior in Fig. \ref{figVnnnGrCCnk}(c). The trion condensate remains stable over a broad range of interaction strengths, specifically for $-5.8\leq U_{bf} \leq -1.2$. Beyond this regime, the system becomes unstable and undergoes phase separation (PS), where clustered particles are randomly distributed without a well-defined central charge. 
Additionally, the trion condensate is favored when $t_b<t_f$, corresponding to heavy bosons and light fermions. For a fixed $U_{bf}=-V_{bf}=-5.0$, the composite trion phase remains robust within $0.4\leq t_b/t_f\leq 4.4$, as shown in Fig. \ref{figccVfftb}.
%

\section{Discussions}
Our study unravels a general mechanism for forming a condensate of composite bosonic trions. Rather than relying on complex three-body interactions, our approach leverages the combination of preformed fermion pairs with bosons via simple two-body boson–fermion interactions.
For the pair–hopping model, recent advances in Floquet engineering enable the simulation of pair hopping by enhancing pair tunneling while simultaneously suppressing single-particle tunneling \cite{exp2024pairTunneling,Eckardt2017RevMod,Bloch2007pairTunneling}. Furthermore, Feshbach resonances provide a convenient means to tune boson–fermion attractions \cite{Tools,doi:10.1126/science.aal3837,RevModPhys.80.885}, making the implementation of the pair–hopping model experimentally promising.
In parallel, the extended Bose–Fermi Hubbard model—describing dipolar bosons and fermions confined in a one-dimensional optical lattice—offers an alternative route for realizing composite trion condensates. 
Combined with Feshbach resonance techniques, both short-range onsite interactions and long-range dipole–dipole interactions (including first- and second-neighbor density–density couplings) can be precisely controlled \cite{Zoller2012dipolarReview,Tools,doi:10.1126/science.aal3837}.
Notably, the composite trion phase occupies a substantial portion of the phase diagrams for both models, further enhancing its experimental feasibility. 
The formation of composite trions can be identified by the radio-frequency (rf) dissociation spectrum, and their binding energy can be measured based on the rf loss spectrum \cite{Pan2022CreationAtomMoleculeNaKK,Pan2022EvidenceAtomMoleculeNaKK,Pan2024PhotoassociationNaKK}.
Furthermore, the condensation of bosonic trions is expected to manifest in measurements of the pair momentum distribution \cite{exp2015PairCondensation,multimerDetect2007,ferimionicTrimer2010}. Meanwhile, the suppression of superconducting correlations of fermions can be detected using interferometric techniques \cite{corrSCexp2005}.

In conclusion, we have demonstrated that a quasi–long–range coherent ground state of composite bosonic trions, characterized by a central charge $c=1$, emerges robustly in both the pair–hopping model and the extended Bose–Fermi Hubbard model. This phase is stabilized by a combination of negative binding energy, gapped single–particle excitations, and suppressed correlations of paired fermions, which together underpin the quasi–condensation of the bosonic trions and illuminate the underlying pairing mechanism between bosons and fermions.
Our results also confirm the existence of the trion phase in multi-leg ladder systems beyond the 1D chain geometry (see \ref{AppendixC}). 
While our calculations have been conducted for simplicity on spinless fermions and bosons, we believe our results can be generalized to spinful systems, resulting in a richer phase diagram. 
Overall, our findings shed light on the three-body pairing mechanism in Bose-Fermi mixtures and open avenues for future experimental and theoretical investigations into intriguing pairing phases.

\section*{Acknowledgements}
We thank Ting-Kuo Lee and Kai Chang for their invaluable suggestions. This work is supported by the National Key Research and Development Program of China Grant No. 2022YFA1404204, and the National Natural Science Foundation of China Grant No. 12274086.

\setcounter{section}{0}
\renewcommand{\thesection}{Appendix \Alph{section}} 
\renewcommand{\thesubsection}{\arabic{subsection}}
\renewcommand{\thefigure}{S\arabic{figure}}
\setcounter{figure}{0} 
\renewcommand{\thetable}{S\arabic{table}}

\section{Detailed numerical setup}\label{AppendixA}

The numerical calculations employed the state-of-the-art Density Matrix Renormalization Group (DMRG) method, as referenced in \cite{PhysRevLett.69.2863, ITensor}.
Our simulations studied system sizes of up to $L=160$ with open boundary conditions (OBC) and $L=96$ with periodic boundary conditions (PBC). The computations involved approximately 150 sweeps, retaining $m=1500$ states for OBC with a typical truncation error $\epsilon\sim 10^{-9}$, and $m=3000$ states for PBC with $\epsilon\sim 10^{-8}$.
Due to the substantial onsite repulsion ($U_{bb}=8.0$) and the relatively low boson density ($\rho_b=N_b/L=1/8$), it was adequate to set the maximum boson occupation number per site ($n^{cut}_b$) at 2. Notably, results for $n^{cut}_b \geq 1$ were comparable.

\section{Results for the Bose–Fermi model incorporating fermion pair hopping} \label{AppendixB}

\begin{figure}
\centering
\includegraphics[scale=0.65,trim=0 0 0 0,clip]{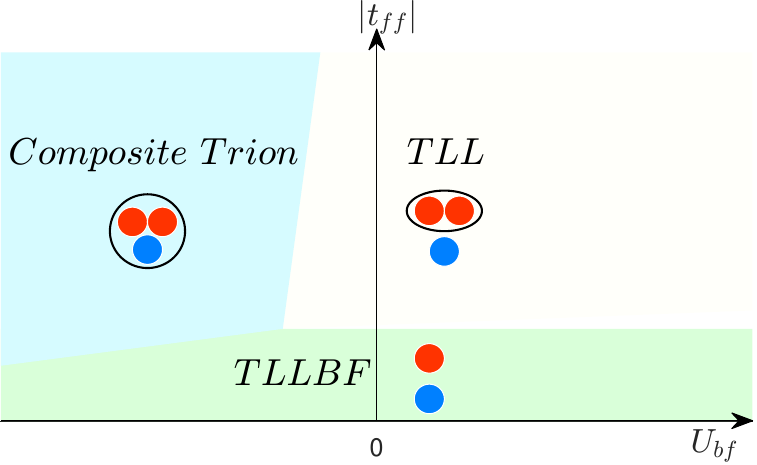}
\caption{\label{figS_PD_tff_Ubf} 
Schematic phase diagram for the Bose–Fermi model incorporating fermion pair hopping in the $t_{ff}$-$ U_{bf}$ parameter space for $\rho_f=2\rho_b=1/4$. 
The composite trion and TLL phases are the same as those described in the main text. "TLL" denotes a two-component Luttinger liquid of bosons and paired fermions, whereas "TLLBF" denotes a two-component Luttinger liquid of bosons and unpaired fermions.
Solid red and blue circles represent fermions and bosons, respectively. 
Particles enclosed within black circles and ellipses indicate paired configurations. Negative values of $t_{ff}$ yield similar results to those for positive $t_{ff}$. 
}
\end{figure}

\begin{figure}[htb]
\centering
\includegraphics[scale=0.7,trim=0 0 0 0,clip]{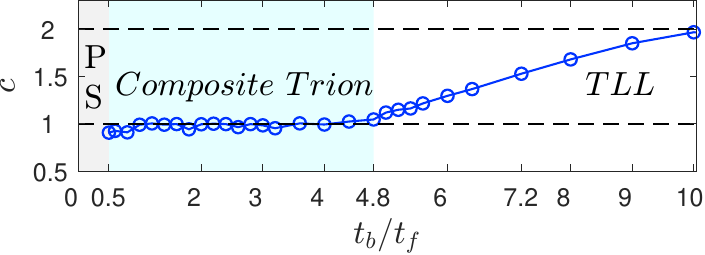}
\caption{\label{figScctfftb}Central charge $c$ as a function of $t_b/t_f$ in the $L=160$ system with OBC. Within the composite trion phase, $c=1$ for $0.5\leq t_b/t_f\leq 4.8$. "PS" denotes phase separation for $t_b/t_f<0.5$, characterized by randomly distributed particle clusters lacking a well-defined central charge. The parameters are set as $t_f=1,t_{ff}=3.0,U_{bf}=-2.8,U_{bb}=8.0$, and $\rho_f=2\rho_b=1/4$.
}
\end{figure}

\subsection{Schematic phase diagram  as a function of $t_{ff}$ and $U_{bf}$}

We present a qualitative ground-state phase diagram as a function of $t_{ff}$ and $U_{bf}$ for fermion and boson densities $\rho_f=2\rho_b=1/4$, as shown in Fig.~\ref{figS_PD_tff_Ubf}. Negative values of $t_{ff}$ yield results similar to those for positive $t_{ff}$.

For small fermion pair-hopping amplitudes $t_{ff}$, the system approaches the limit of the standard Bose–Fermi Hubbard model \cite{bfmPhaseDiagram2006}. In this regime, the ground state is a two-component Luttinger liquid of bosons and unbound fermions, exhibiting a central charge of $c=2$. We refer to this as the “TLLBF” phase, where the single-particle correlation functions of both fermions and bosons display dominant quasi-long-range order. Due to the fermion density being twice that of the bosons, the TLLBF phase persists even at large boson–fermion interactions $U_{bf}$. In contrast, when the densities of the two species are equal and $U_{bf}$ is large, the ground state becomes a single-component Luttinger liquid of composite fermions formed by binding one fermion with one boson \cite{bfmPhaseDiagram2006}.
Excessively strong $U_{bf}$ at $\rho_f=2\rho_b$ drives the system into a phase-separated state, where particle clusters distribute in distinct regions of the real-space lattice. This behavior can be intuitively understood as follows: at $\rho_f=2\rho_b$,  unbound fermions and composite fermions coexist in the system, and the strong scattering of composite fermions by the remaining unbound fermions leads to phase separation\cite{QiSong2024CSF}.
This regime lies beyond the scope of the schematic phase diagram. For instance, phase separation occurs at $U_{bf}=-28$ for $t_{ff}=-0.2$ with parameters $t_f=t_b=1$ and $U_{bb}=8$.

\begin{figure}
\centering
\includegraphics[scale=0.75,trim=4 0 1 0,clip]{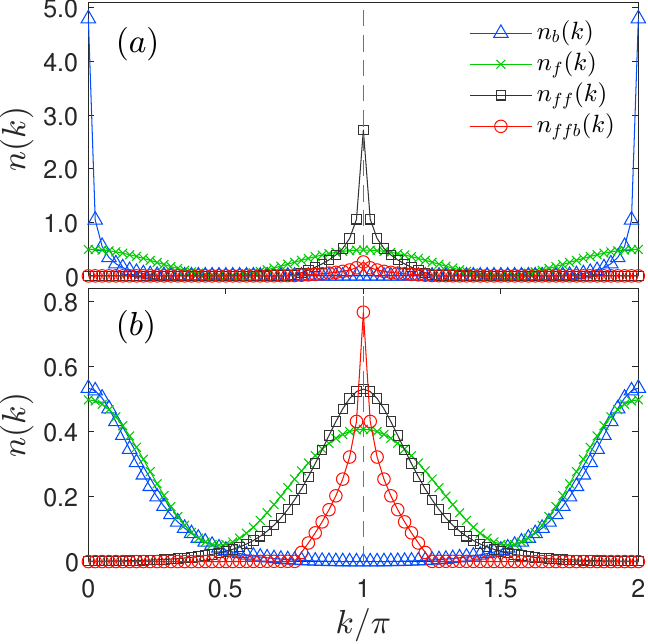}
\caption{\label{fig2_nk_tff+}Momentum distribution functions $n(k)$ for $t_{ff}=+3.0,\rho_f=2\rho_b=1/4$ in a system of length $L=80$ with PBC. Panel (a) displays the results for the TLL phase at $U_{bf}=-0.3$, while panel (b) shows the results for the composite trion phase at $U_{bf}=-6.0$. The grey dashed lines guide the eyes at $k=\pi$.
}
\end{figure}

For large $t_{ff}$ and small attractive $U_{bf}$, the system resides in a two-component Luttinger liquid phase consisting of bosons and paired fermions, denoted as the TLL phase in the main text. This TLL phase also appears in the repulsive $U_{bf}$ regime.
When $U_{bf}=0$ and $t_{ff}$ is positive, the system exhibits a coexistence phase comprising neighboring paired fermions in a sea of unpaired fermions \cite{TwoFluid2021}. However, as $U_{bf}$ increases, this coexistence region rapidly narrows and eventually disappears. Furthermore, since this phase has negligible influence on the composite trion phase and occupies only a very small parameter region, it is not shown in the schematic phase diagram.

Additionally, for large $t_{ff}$ and strongly attractive $U_{bf}$, the composite trion phase occupies a broad region of the phase diagram. In this phase, only the composite bosonic trions exhibit dominant quasi-long-range order.




\subsection{Influence of magnitude of $t_b/t_f$}

The appearance of the trion condensate is facilitated when $t_b<t_f$, but it also remains stable over a wide range of $t_b\geq t_f$. For example, when $U_{bf}= -2.8t_f$, we observe that for $0.5\leq t_b/t_f \leq 4.8$, the system consistently remains in the composite trion phase, as shown in Fig. \ref{figScctfftb}. This demonstrates that the hopping amplitudes $t_b$ and $t_f$ do not require highly precise tuning, and the trion condensate can emerge over a broad range of values. For too small $t_b/t_f$, the system becomes unstable to phase separated phase (PS), where clustered particles are randomly distributed without a well-defined central charge. 


\begin{figure}
\centering
\includegraphics[scale=0.62,trim=0 0 0 0,clip]{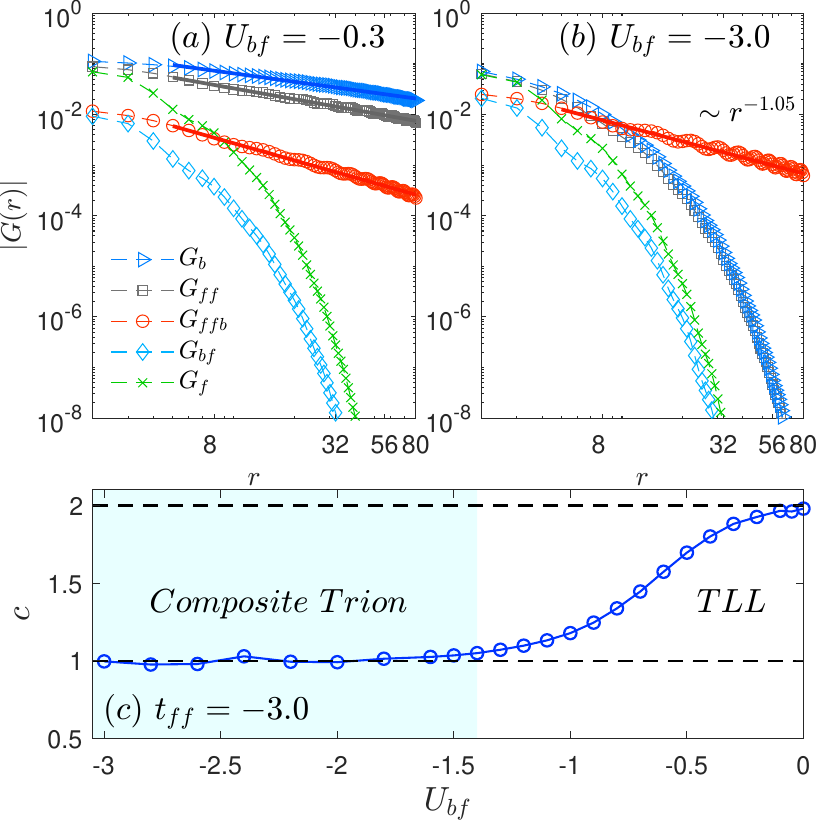}
\caption{\label{figS1_Gr_cc} Correlation functions for (a) the TLL phase at $U_{bf}=-0.3$ and (b) the composite trion phase at $U_{bf}=-3.0$ in a 1D system of length $L=160$ with OBC. The parameters $t_{ff}=-3.0,\rho_f=2\rho_b=1/4$ are fixed and $r$ represents the distance from the reference site at $L/4$. Both (a) and (b) use a double-logarithmic scale. The solid lines denote power-law fitting to $|G(r)|\sim r^{-\alpha_q}$, where $\alpha_b\simeq 0.55, \alpha_{ff}\simeq 0.71, \alpha_{ffb}\simeq 1.13$ in (a), and $\alpha_{fbf}\simeq 1.05$ in (b).  (c) The central charge is obtained by fitting the von Neumann entropy as a function of $U_{bf}$.  In the composite trion phase, $c=1$ when $U_{bf}\leq -1.4$.
}
\end{figure}

\begin{figure*}
\centering
\includegraphics[scale=0.7,trim=4 0 0 0,clip]{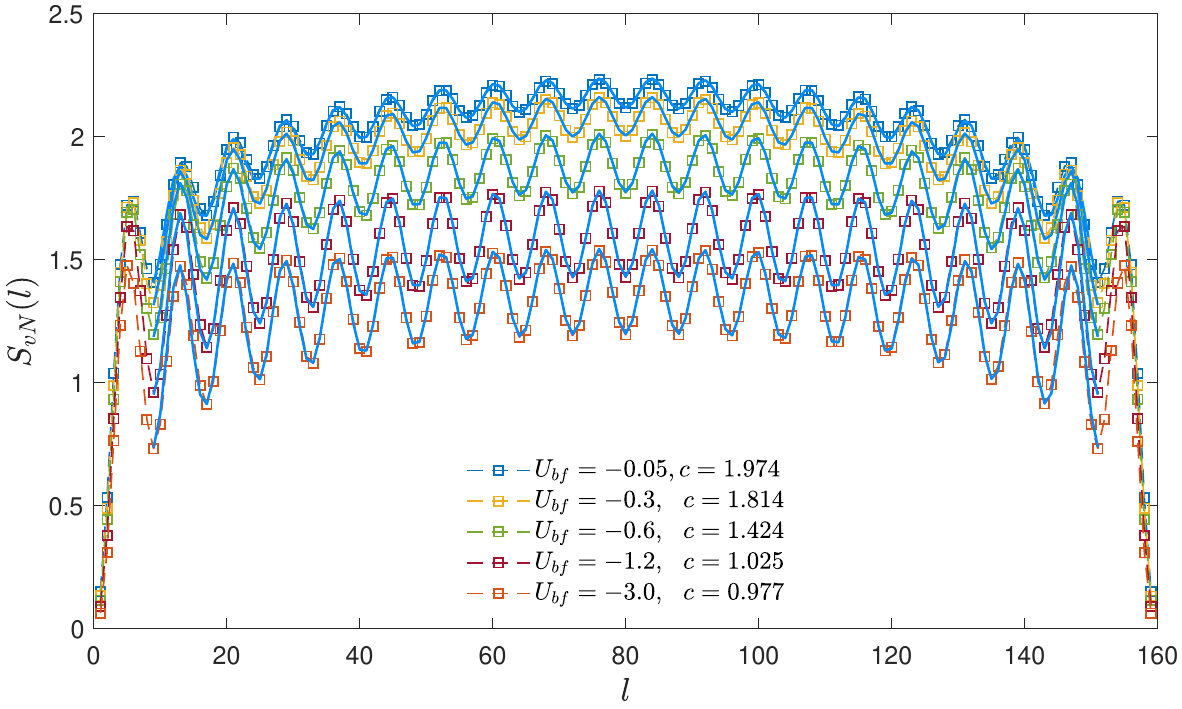}
\caption{\label{figS_SvN_tff+3.0} Fitting of the central charge $c$ from the von Neumann entanglement entropy $S_{vN}$ in the $L=160$ system with OBC at $t_{ff}=3.0$. Here, $l$ represents the subsystem length. The open squares depict numerical data, while the solid lines are fitted using Eq. (\ref{eq_SvN}).
}
\end{figure*}

\subsection{Results for Negative $t_{ff}=-3.0$}
In the limit of paired fermions ($|t_{ff}|\gg t_f$), the sign of $t_{ff}$ primarily affects the momenta where fermion pairs quasi-condensate, specifically at $k=0$ for $t_{ff}<0$ and $k=\pi$ for $t_{ff}>0$). Regardless of the sign of $t_{ff}$, bosons distribute predominantly around $k=0$ in momentum space. 
Since paired fermions are constituent parts of composite trions, the condensation momenta of the composite trions are either $k=0$ for $t_{ff}<0$ (see Fig. 4 in the main text) or $k=\pi$ for $t_{ff}>0$ (see Fig. \ref{fig2_nk_tff+}(b)). 
The evolution of the correlation functions and central charge for $t_{ff}=-3.0$ display similar behaviors as those for $t_{ff}=+3.0$, which can be seen in Fig. \ref{figS1_Gr_cc} here and Fig. 2 in the main text.
For $t_{ff}=-3.0$, the system fully transmutes into the composite trion phase with a central charge equal to 1 when $U_{bf}\leq -1.4$, which slightly differs from the case for $t_{ff}=+3.0$.

\begin{figure*}
\centering
\includegraphics[scale=0.7,trim=0 0 1 0,clip]{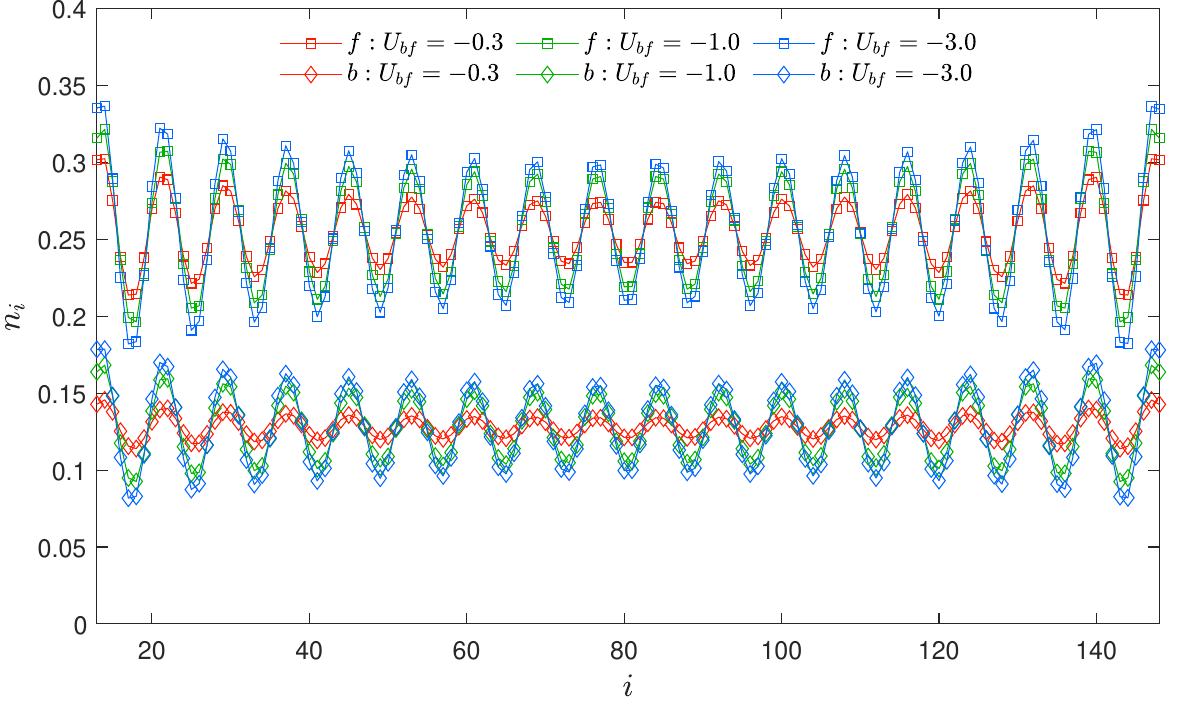}
\caption{\label{figS_ni}Charge density profiles $n_i$ at site $i$ for $t_{ff}=3.0,\rho_f=2\rho_b=1/4$ at various $U_{bf}$ values in the $L=160$ system with OBC. "f" and "b" denote fermions and bosons, respectively.
}
\end{figure*}

\subsection{Entanglement entropy and central charge} 
The low energy physics of both the TLL phase and the composite trion phase can be described by a (1+1)-dimensional conformal field theory. The central charge of this conformal field theory is determined by calculating the von Neumann entanglement entropy, defined as $S_{vN}=-$Tr [$\rho_l$ ln$ \rho_l]$, where $\rho_l$ represents the reduced density matrix of a subsystem with length $l$. For open systems,  the central charge $c$ can be deduced using the relation  \cite{TwoFluid2021,fpairingPhysRevResearch2021,SvN2009spin,Calabrese_2009,Calabrese_2004}
\begin{eqnarray} \label{eq_SvN}
S_{vN}(l)=&&\frac{c}{6}ln(\frac{2L}{\pi}sin(\frac{\pi l}{L})) \nonumber \\
&&+B(\langle c^{\dagger}_lc_{l+1} \rangle +H.c.)+S_0 ,
\end{eqnarray}
where $S_0$ and $B$ are model-dependent fitting parameters, and $L$ denotes the system size. Due to the presence of additional oscillations in $S_{vN}(l)$, the local fermion hopping terms $B(\langle c^{\dagger}_lc_{l+1} \rangle +H.c.)$ are incorporated, with $B$ depending on the fitting \cite{TwoFluid2021,fpairingPhysRevResearch2021,SvN2009spin}. In the TLL phase, $c=2$ corresponds to the two gapless modes: the Luttinger liquid of paired fermions and that of bosons. In the composite trion phase, $c=1$ corresponds to a single bosonic gapless mode of the Luttinger liquid of composite trions. Fig. \ref{figS_SvN_tff+3.0} shows several typical fits of the central charge $c$ from the von Neumann entanglement entropy $S_{vN}$.

\begin{figure}
\centering
\includegraphics[scale=0.65,trim=0 0 1 0,clip]{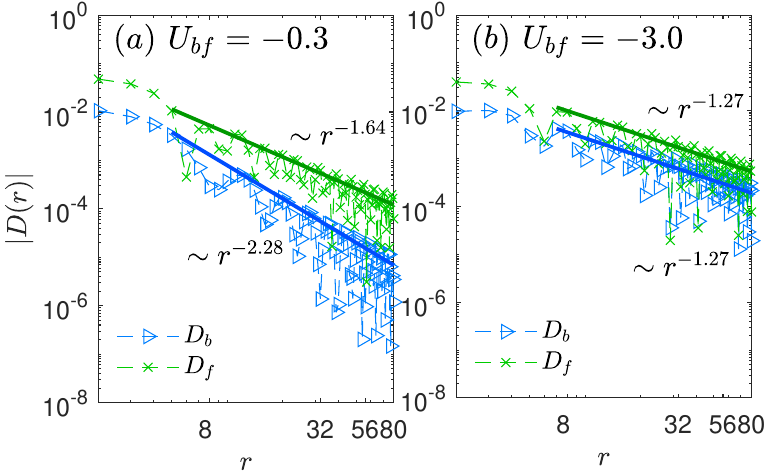}
\caption{\label{figS_Dij}Density-density correlations with $t_{ff}=3.0$ for (a) the TLL phase at $U_{bf}=-0.3$ and (b) the composite trion phase at $U_{bf}=-3.0$ in a 1D system of length $L=160$ with OBC. $\rho_f=2\rho_b=1/4$ are fixed, and $r$ represents the distance from the reference site at $L/4$. A double-logarithmic scale is used in both (a) and (b). Solid lines denote power-law fitting to $|G(r)|\sim r^{-\alpha_q}$ where $\alpha_b\simeq 2.28,\alpha_f\simeq 1.64$ in (a) and $\alpha_{f}\simeq 1.27,\alpha_{b}\simeq 1.27$ in (b).   
}
\end{figure}
\begin{figure}
\centering
\includegraphics[scale=0.7,trim=4 0 1 0,clip]{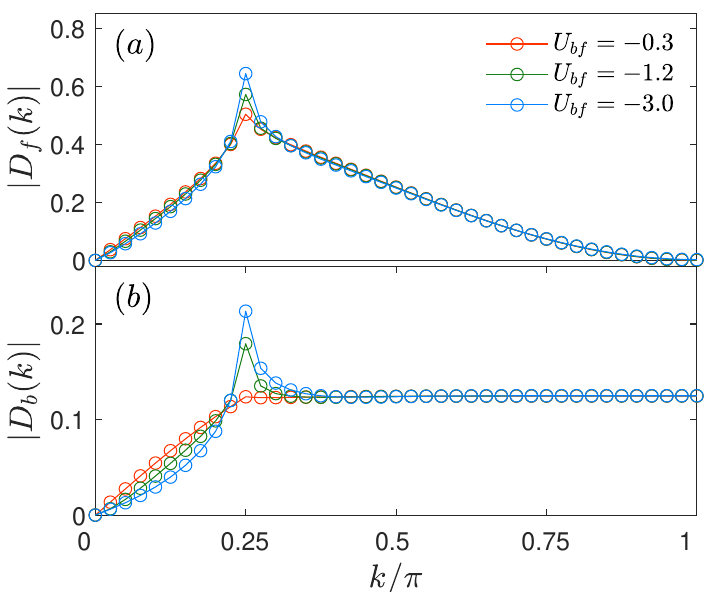}
\caption{\label{figS_Dk} Structure factors of density correlation functions for a periodic system with $L=80$ for (a) fermions and (b) bosons at various $U_{bf} $ values.
}
\end{figure}

\subsection{Charge density waves} 
In 1D chains with open boundary conditions, Friedel oscillations can be observed in the charge density profiles\cite{tJMoreno2011}. A $2k_F\ (k_F=\pi\rho_f)$ charge-density-wave (CDW) order persists in both the TLL phase and the composite trion phase, as shown in Fig. \ref{figS_ni}. The $2k_F$ periodicity of CDW, rather than $k_F$, arises due to bound fermion pairs at $|t_{ff}|=3.0$.  Bosons exhibit the same oscillatory behavior as fermions due to the onsite attractive interaction $U_{bf}$. At small $|U_{bf}|$ values, the density-density correlation functions of both species exhibit algebraic decay with power exponents close to $2$, indicating weak CDW order (see Fig. \ref{figS_Dij}). As $U_{bf}$ increases, the amplitude of CDW is enhanced and the density-density correlations decay slower with power exponents approaching $1$, indicating a quasi-long-range CDW order. The decay exponents of both species are nearly identical, suggesting that bosons are locked to fermions in the strong coupling regime.

To further characterize the composite trion phase, we calculate the charge-density correlation functions $D_q(r)=\langle n^q_{i}n^q_{j}\rangle-\langle n^q_{i}\rangle \langle n^q_{j}\rangle$ for $q=f,b$. Based on the bosonization prediction for 1D systems\cite{fpairingPhysRevResearch2021},  the lowest order contributions to $D_q(r)$ have the form 
\begin{eqnarray} \label{eq_Dr}
D_q(r)= \frac{A_1}{r^2} + A_2\frac{cos(2\pi \rho r)}{r^{2K_q}},
\end{eqnarray}
where $A_1, A_2$ are model-dependent amplitudes, and $K_q$ is the Luttinger parameter. In both the TLL phase and the composite trion phase, fermions are always in a paired configuration, hence $\rho=\rho_f/2=\rho_b$ for $D_f(r)$. Upon Fourier transformation, the density structure factor $D_{f}(k)$ display a $2k_{c}$ cusp with $k_{c}=\pi\rho$, signaling a $2k_{c}$ charge-density wave. Additionally, for bosons' $D_b(k)$, the $2k_{c}$ cusp is only present in the composite trion phase and absent in the TLL phase. Furthermore, in Fig. \ref{figS_Dk}(b), the $2k_{c}$ cusp becomes more prominent as the absolute value of $U_{bf}$ increases, indicating an enhanced tendency to form composite trions.

\subsection{Results for fermions at half filling: $\rho_f=2\rho_b=\frac{1}{2}$} 

\begin{figure}
\centering
\includegraphics[scale=0.62,trim=0 0 0 0,clip]{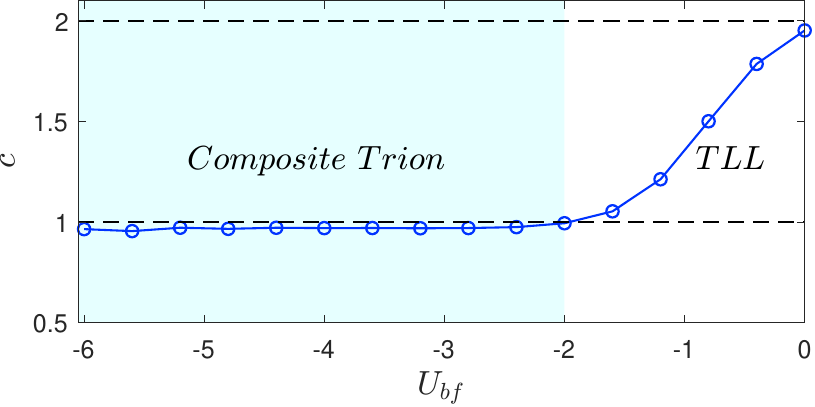}
\caption{\label{figS_cc_rho0.5_Ubf-3}The central charge for a system with $\rho_f=2\rho_b=1/2$ at $t_{ff}=3.0$ in a 1D system of length $L=160$ with OBC.  In the composite trion phase, $c=1$ when $U_{bf}\leq -2.0$.
}
\end{figure}

The results obtained at another typical particle density, $\rho_f=2\rho_b=\frac{1}{2}$, are similar to those at $\rho_f=2\rho_b=\frac{1}{4}$, except for a larger value of $U_{bf}$ where the system fully transitions into the composite trion phase with a central charge of 1. For $\rho_f=2\rho_b=\frac{1}{2}$, the central charge $c=1$ when $U_{bf}\leq -2.0$, whereas for $\rho_f=2\rho_b=\frac{1}{4}$, $c=1$ when $U_{bf}\leq -1.2$. The gradual changes in the central charge as $U_{bf}$ varies are illustrated in Fig. \ref{figS_cc_rho0.5_Ubf-3}. The fermion-boson attraction $U_{bf}$ drives the crossover from the TLL phase with $c=2$ to the composite trion phase with $c=1$.

\section{Stability of composite trions in ladder systems}\label{AppendixC} 

\begin{figure}
\centering
\includegraphics[scale=0.65,trim=4 0 0 0,clip]{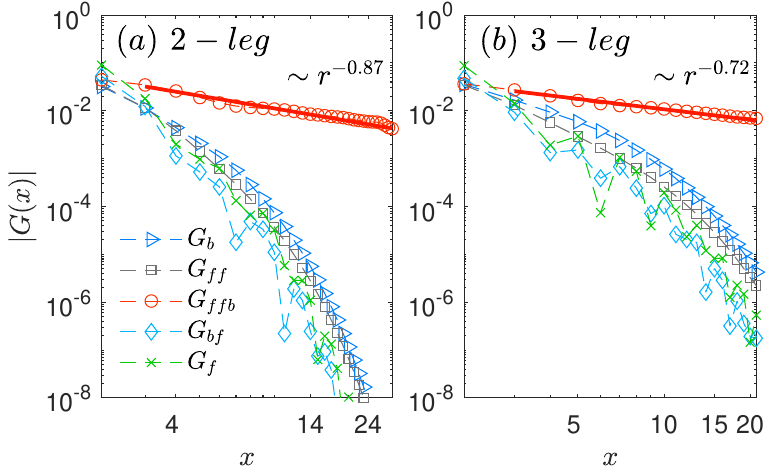}
\caption{\label{fig_lad}Correlation functions in the composite trion phase at $t_{ff}=3.0, U_{bf}=-12.0,U_{bb}=8.0$ for (a) 2-leg ladder with $L_x=48$ and (b) 3-leg cylinder with $L_x=32$. $x$ is the distance from the reference site $L_x/4$ in the $\hat{x}-$direction. Double-logarithmic scale is used. The solid lines denote power-law fitting to $|G(r)|\sim r^{-\alpha_{ffb}}$.  
}
\end{figure}

The composite trion phase is not confined to 1D chain systems but can also stabilize in multi-leg ladder configurations. The Hamiltonian is designed to allow individual fermions and bosons to hop in both the $\hat{x}-$ and $\hat{y}-$directions, while the paired fermion hopping terms are restricted to the $\hat{x}-$direction. The system adopts a cylinder geometry with periodic and open boundary conditions in the $\hat{y}$ and $\hat{x}$ directions. Bosons interact with fermions through a local density-density attraction $U_{bf}$. Correlation functions are averaged over the number of legs $L_y$ as 
$G_O(x)=\frac{1}{L_y}\sum^{L_y}_{y=1}\langle O^{\dagger}(x_0,y)O(x_0+x,y) \rangle$ with the reference site at $x_0\sim L_x/4$. Fig. \ref{fig_lad} demonstrates that, among various correlations, only the composite trion correlation functions decay the slowest, following a power-law manner.

\section{composite fermion pairing phase at $N_f=N_b$}\label{AppendixD} 
\begin{figure}
\centering
\includegraphics[scale=0.65,trim=2 1 1 0,clip]{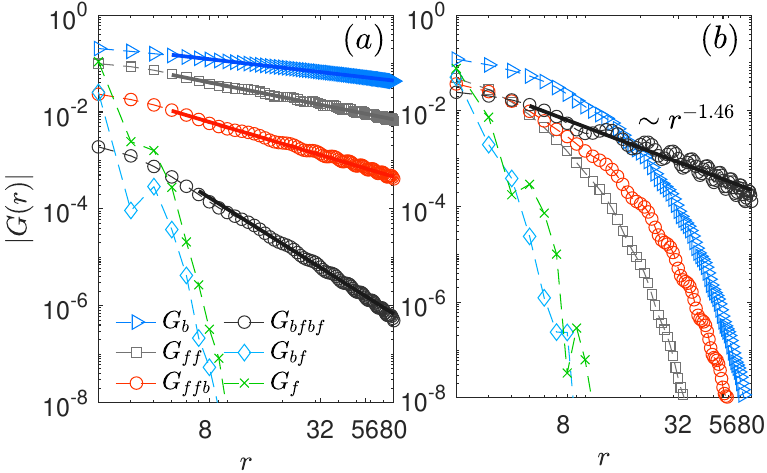}
\caption{\label{figS_Gr_Nf-Nb} Correlation functions with $N_f=N_b$ 
for (a) the TLL phase at $U_{bf}=-0.4$ and (b) the composite fermion pairing (CFP) phase at $U_{bf}=-4.0$ in a 1Dsystem of length $L=160$ with OBC. Here, $t_{ff}=3.0,\rho_f=\rho_b=1/4$ are fixed, and $r$ represents the distance from the reference site at $L/4$. A double-logarithmic scale is used in both (a) and (b). The solid lines epresent power-law fitting to $|G(r)|\sim r^{-\alpha_q}$ where $\alpha_b\simeq 0.44, \alpha_{ff}\simeq 0.76, \alpha_{ffb}\simeq 1.13$, $\alpha_{bfbf}\simeq 2.4$ in (a) and $\alpha_{bfbf}\simeq 1.46$ in (b). 
}
\end{figure}
When the densities of fermions and bosons are equal, a composite fermion pairing phase emerges. A composite fermion consists of one fermion and one boson. At large $|U_{bf}|$ values, only the correlation function of paired composite fermions, given by 
\begin{eqnarray} \label{eq_Gbfbf}
G_{bfbf}(r)= \langle c^{\dagger}_ib^{\dagger}_ic^{\dagger}_{i+1}b^{\dagger}_{i+1}c_{j}b_{j}c_{j+1}b_{j+1} \rangle,r=|i-j| ,
\end{eqnarray}
demonstrates power-law decay behavior with exponents $\alpha_{bfbf}<2$, as shown in Fig. \ref{figS_Gr_Nf-Nb}. Consequently, the dominant QLRO is characterized by the pairing of composite fermions. The pairing phase of composite fermions, consisting of one fermion and one bosonic hole, has been discussed earlier in the context of polarons\cite{PhysRevA.87.021603,PhysRevLett.93.120404,PhysRevA.75.013612}. Here, polarons refer to atoms of one species dressed by screening clouds formed by atoms of another species.
The identification of this composite-fermion-paired superconducting phase further enriches the landscape of unconventional many-body pairing states in Bose–Fermi systems.

\bibliography{trionPRA}

\begin{thebibliography}{77}%
\makeatletter
\providecommand \@ifxundefined [1]{%
 \@ifx{#1\undefined}
}%
\providecommand \@ifnum [1]{%
 \ifnum #1\expandafter \@firstoftwo
 \else \expandafter \@secondoftwo
 \fi
}%
\providecommand \@ifx [1]{%
 \ifx #1\expandafter \@firstoftwo
 \else \expandafter \@secondoftwo
 \fi
}%
\providecommand \natexlab [1]{#1}%
\providecommand \enquote  [1]{``#1''}%
\providecommand \bibnamefont  [1]{#1}%
\providecommand \bibfnamefont [1]{#1}%
\providecommand \citenamefont [1]{#1}%
\providecommand \href@noop [0]{\@secondoftwo}%
\providecommand \href [0]{\begingroup \@sanitize@url \@href}%
\providecommand \@href[1]{\@@startlink{#1}\@@href}%
\providecommand \@@href[1]{\endgroup#1\@@endlink}%
\providecommand \@sanitize@url [0]{\catcode `\\12\catcode `\$12\catcode `\&12\catcode `\#12\catcode `\^12\catcode `\_12\catcode `\%12\relax}%
\providecommand \@@startlink[1]{}%
\providecommand \@@endlink[0]{}%
\providecommand \url  [0]{\begingroup\@sanitize@url \@url }%
\providecommand \@url [1]{\endgroup\@href {#1}{\urlprefix }}%
\providecommand \urlprefix  [0]{URL }%
\providecommand \Eprint [0]{\href }%
\providecommand \doibase [0]{http://dx.doi.org/}%
\providecommand \selectlanguage [0]{\@gobble}%
\providecommand \bibinfo  [0]{\@secondoftwo}%
\providecommand \bibfield  [0]{\@secondoftwo}%
\providecommand \translation [1]{[#1]}%
\providecommand \BibitemOpen [0]{}%
\providecommand \bibitemStop [0]{}%
\providecommand \bibitemNoStop [0]{.\EOS\space}%
\providecommand \EOS [0]{\spacefactor3000\relax}%
\providecommand \BibitemShut  [1]{\csname bibitem#1\endcsname}%
\let\auto@bib@innerbib\@empty
\bibitem [{\citenamefont {Hammer}\ and\ \citenamefont {Platter}(2010)}]{hammer2010efimov}%
  \BibitemOpen
  \bibfield  {author} {\bibinfo {author} {\bibfnamefont {H.-W.}\ \bibnamefont {Hammer}}\ and\ \bibinfo {author} {\bibfnamefont {L.}~\bibnamefont {Platter}},\ }\href {\doibase https://doi.org/10.1146/annurev.nucl.012809.104439} {\bibfield  {journal} {\bibinfo  {journal} {Annual Review of Nuclear and Particle Science}\ }\textbf {\bibinfo {volume} {60}},\ \bibinfo {pages} {207} (\bibinfo {year} {2010})}\BibitemShut {NoStop}%
\bibitem [{\citenamefont {Guan}\ \emph {et~al.}(2013)\citenamefont {Guan}, \citenamefont {Batchelor},\ and\ \citenamefont {Lee}}]{RevModPhys.85.1633}%
  \BibitemOpen
  \bibfield  {author} {\bibinfo {author} {\bibfnamefont {X.-W.}\ \bibnamefont {Guan}}, \bibinfo {author} {\bibfnamefont {M.~T.}\ \bibnamefont {Batchelor}}, \ and\ \bibinfo {author} {\bibfnamefont {C.}~\bibnamefont {Lee}},\ }\href {\doibase 10.1103/RevModPhys.85.1633} {\bibfield  {journal} {\bibinfo  {journal} {Rev. Mod. Phys.}\ }\textbf {\bibinfo {volume} {85}},\ \bibinfo {pages} {1633} (\bibinfo {year} {2013})}\BibitemShut {NoStop}%
\bibitem [{\citenamefont {Naidon}\ and\ \citenamefont {Endo}(2017)}]{Naidon_2017}%
  \BibitemOpen
  \bibfield  {author} {\bibinfo {author} {\bibfnamefont {P.}~\bibnamefont {Naidon}}\ and\ \bibinfo {author} {\bibfnamefont {S.}~\bibnamefont {Endo}},\ }\href {\doibase 10.1088/1361-6633/aa50e8} {\bibfield  {journal} {\bibinfo  {journal} {Reports on Progress in Physics}\ }\textbf {\bibinfo {volume} {80}},\ \bibinfo {pages} {056001} (\bibinfo {year} {2017})}\BibitemShut {NoStop}%
\bibitem [{\citenamefont {Peskin}(2018)}]{peskin2018introduction}%
  \BibitemOpen
  \bibfield  {author} {\bibinfo {author} {\bibfnamefont {M.~E.}\ \bibnamefont {Peskin}},\ }\href@noop {} {\emph {\bibinfo {title} {An Introduction to quantum field theory}}}\ (\bibinfo  {publisher} {CRC press},\ \bibinfo {year} {2018})\BibitemShut {NoStop}%
\bibitem [{\citenamefont {Efimov}(1973)}]{Efimov1973}%
  \BibitemOpen
  \bibfield  {author} {\bibinfo {author} {\bibfnamefont {V.}~\bibnamefont {Efimov}},\ }\href {\doibase https://doi.org/10.1016/0375-9474(73)90510-1} {\bibfield  {journal} {\bibinfo  {journal} {Nuclear Physics A}\ }\textbf {\bibinfo {volume} {210}},\ \bibinfo {pages} {157} (\bibinfo {year} {1973})}\BibitemShut {NoStop}%
\bibitem [{\citenamefont {Kolganova}\ \emph {et~al.}(2017)\citenamefont {Kolganova}, \citenamefont {Motovilov},\ and\ \citenamefont {Sandhas}}]{4HeTrimerFewBody2017}%
  \BibitemOpen
  \bibfield  {author} {\bibinfo {author} {\bibfnamefont {E.}~\bibnamefont {Kolganova}}, \bibinfo {author} {\bibfnamefont {A.}~\bibnamefont {Motovilov}}, \ and\ \bibinfo {author} {\bibfnamefont {W.}~\bibnamefont {Sandhas}},\ }\href {\doibase https://doi.org/10.1007/s00601-016-1181-2} {\bibfield  {journal} {\bibinfo  {journal} {Few-Body Systems}\ }\textbf {\bibinfo {volume} {58}},\ \bibinfo {pages} {35} (\bibinfo {year} {2017})}\BibitemShut {NoStop}%
\bibitem [{\citenamefont {Kunitski}\ \emph {et~al.}(2015)\citenamefont {Kunitski}, \citenamefont {Zeller}, \citenamefont {Voigtsberger}, \citenamefont {Kalinin}, \citenamefont {Schmidt}, \citenamefont {Sch{\"o}ffler}, \citenamefont {Czasch}, \citenamefont {Sch{\"o}llkopf}, \citenamefont {Grisenti}, \citenamefont {Jahnke} \emph {et~al.}}]{4He3Efimov2015}%
  \BibitemOpen
  \bibfield  {author} {\bibinfo {author} {\bibfnamefont {M.}~\bibnamefont {Kunitski}}, \bibinfo {author} {\bibfnamefont {S.}~\bibnamefont {Zeller}}, \bibinfo {author} {\bibfnamefont {J.}~\bibnamefont {Voigtsberger}}, \bibinfo {author} {\bibfnamefont {A.}~\bibnamefont {Kalinin}}, \bibinfo {author} {\bibfnamefont {L.~P.~H.}\ \bibnamefont {Schmidt}}, \bibinfo {author} {\bibfnamefont {M.}~\bibnamefont {Sch{\"o}ffler}}, \bibinfo {author} {\bibfnamefont {A.}~\bibnamefont {Czasch}}, \bibinfo {author} {\bibfnamefont {W.}~\bibnamefont {Sch{\"o}llkopf}}, \bibinfo {author} {\bibfnamefont {R.~E.}\ \bibnamefont {Grisenti}}, \bibinfo {author} {\bibfnamefont {T.}~\bibnamefont {Jahnke}},  \emph {et~al.},\ }\href {\doibase https://doi.org/10.1126/science.aaa5601} {\bibfield  {journal} {\bibinfo  {journal} {Science}\ }\textbf {\bibinfo {volume} {348}},\ \bibinfo {pages} {551} (\bibinfo {year} {2015})}\BibitemShut {NoStop}%
\bibitem [{\citenamefont {Williams}\ \emph {et~al.}(2009)\citenamefont {Williams}, \citenamefont {Hazlett}, \citenamefont {Huckans}, \citenamefont {Stites}, \citenamefont {Zhang},\ and\ \citenamefont {O'Hara}}]{fermionicTrimer2009exp}%
  \BibitemOpen
  \bibfield  {author} {\bibinfo {author} {\bibfnamefont {J.~R.}\ \bibnamefont {Williams}}, \bibinfo {author} {\bibfnamefont {E.~L.}\ \bibnamefont {Hazlett}}, \bibinfo {author} {\bibfnamefont {J.~H.}\ \bibnamefont {Huckans}}, \bibinfo {author} {\bibfnamefont {R.~W.}\ \bibnamefont {Stites}}, \bibinfo {author} {\bibfnamefont {Y.}~\bibnamefont {Zhang}}, \ and\ \bibinfo {author} {\bibfnamefont {K.~M.}\ \bibnamefont {O'Hara}},\ }\href {\doibase 10.1103/PhysRevLett.103.130404} {\bibfield  {journal} {\bibinfo  {journal} {Phys. Rev. Lett.}\ }\textbf {\bibinfo {volume} {103}},\ \bibinfo {pages} {130404} (\bibinfo {year} {2009})}\BibitemShut {NoStop}%
\bibitem [{\citenamefont {Nakajima}\ \emph {et~al.}(2011)\citenamefont {Nakajima}, \citenamefont {Horikoshi}, \citenamefont {Mukaiyama}, \citenamefont {Naidon},\ and\ \citenamefont {Ueda}}]{fermionicTrimer2011exp}%
  \BibitemOpen
  \bibfield  {author} {\bibinfo {author} {\bibfnamefont {S.}~\bibnamefont {Nakajima}}, \bibinfo {author} {\bibfnamefont {M.}~\bibnamefont {Horikoshi}}, \bibinfo {author} {\bibfnamefont {T.}~\bibnamefont {Mukaiyama}}, \bibinfo {author} {\bibfnamefont {P.}~\bibnamefont {Naidon}}, \ and\ \bibinfo {author} {\bibfnamefont {M.}~\bibnamefont {Ueda}},\ }\href {\doibase 10.1103/PhysRevLett.106.143201} {\bibfield  {journal} {\bibinfo  {journal} {Phys. Rev. Lett.}\ }\textbf {\bibinfo {volume} {106}},\ \bibinfo {pages} {143201} (\bibinfo {year} {2011})}\BibitemShut {NoStop}%
\bibitem [{\citenamefont {Shi}\ \emph {et~al.}(2014)\citenamefont {Shi}, \citenamefont {Cui},\ and\ \citenamefont {Zhai}}]{fermionicTrimer2014HuiZhai}%
  \BibitemOpen
  \bibfield  {author} {\bibinfo {author} {\bibfnamefont {Z.-Y.}\ \bibnamefont {Shi}}, \bibinfo {author} {\bibfnamefont {X.}~\bibnamefont {Cui}}, \ and\ \bibinfo {author} {\bibfnamefont {H.}~\bibnamefont {Zhai}},\ }\href {\doibase 10.1103/PhysRevLett.112.013201} {\bibfield  {journal} {\bibinfo  {journal} {Phys. Rev. Lett.}\ }\textbf {\bibinfo {volume} {112}},\ \bibinfo {pages} {013201} (\bibinfo {year} {2014})}\BibitemShut {NoStop}%
\bibitem [{\citenamefont {Orso}\ \emph {et~al.}(2010)\citenamefont {Orso}, \citenamefont {Burovski},\ and\ \citenamefont {Jolicoeur}}]{ferimionicTrimer2010}%
  \BibitemOpen
  \bibfield  {author} {\bibinfo {author} {\bibfnamefont {G.}~\bibnamefont {Orso}}, \bibinfo {author} {\bibfnamefont {E.}~\bibnamefont {Burovski}}, \ and\ \bibinfo {author} {\bibfnamefont {T.}~\bibnamefont {Jolicoeur}},\ }\href {\doibase 10.1103/PhysRevLett.104.065301} {\bibfield  {journal} {\bibinfo  {journal} {Phys. Rev. Lett.}\ }\textbf {\bibinfo {volume} {104}},\ \bibinfo {pages} {065301} (\bibinfo {year} {2010})}\BibitemShut {NoStop}%
\bibitem [{\citenamefont {Schwartz}\ \emph {et~al.}(2021)\citenamefont {Schwartz}, \citenamefont {Shimazaki}, \citenamefont {Kuhlenkamp}, \citenamefont {Watanabe}, \citenamefont {Taniguchi}, \citenamefont {Kroner},\ and\ \citenamefont {Imamo{\u{g}}lu}}]{TMDexp2021}%
  \BibitemOpen
  \bibfield  {author} {\bibinfo {author} {\bibfnamefont {I.}~\bibnamefont {Schwartz}}, \bibinfo {author} {\bibfnamefont {Y.}~\bibnamefont {Shimazaki}}, \bibinfo {author} {\bibfnamefont {C.}~\bibnamefont {Kuhlenkamp}}, \bibinfo {author} {\bibfnamefont {K.}~\bibnamefont {Watanabe}}, \bibinfo {author} {\bibfnamefont {T.}~\bibnamefont {Taniguchi}}, \bibinfo {author} {\bibfnamefont {M.}~\bibnamefont {Kroner}}, \ and\ \bibinfo {author} {\bibfnamefont {A.}~\bibnamefont {Imamo{\u{g}}lu}},\ }\href {https://doi.org/10.1126/science.abj3831} {\bibfield  {journal} {\bibinfo  {journal} {Science}\ }\textbf {\bibinfo {volume} {374}},\ \bibinfo {pages} {336} (\bibinfo {year} {2021})}\BibitemShut {NoStop}%
\bibitem [{\citenamefont {Wagner}\ \emph {et~al.}(2025)\citenamefont {Wagner}, \citenamefont {O\l{}dziejewski}, \citenamefont {Rose}, \citenamefont {K\"oder}, \citenamefont {Kuhlenkamp}, \citenamefont {\ifmmode \dot{I}\else \.{I}\fi{}mamo\ifmmode~\breve{g}\else \u{g}\fi{}lu},\ and\ \citenamefont {Schmidt}}]{TMDtheory2025}%
  \BibitemOpen
  \bibfield  {author} {\bibinfo {author} {\bibfnamefont {M.}~\bibnamefont {Wagner}}, \bibinfo {author} {\bibfnamefont {R.}~\bibnamefont {O\l{}dziejewski}}, \bibinfo {author} {\bibfnamefont {F.}~\bibnamefont {Rose}}, \bibinfo {author} {\bibfnamefont {V.}~\bibnamefont {K\"oder}}, \bibinfo {author} {\bibfnamefont {C.}~\bibnamefont {Kuhlenkamp}}, \bibinfo {author} {\bibfnamefont {A.~m.~c.}\ \bibnamefont {\ifmmode \dot{I}\else \.{I}\fi{}mamo\ifmmode~\breve{g}\else \u{g}\fi{}lu}}, \ and\ \bibinfo {author} {\bibfnamefont {R.}~\bibnamefont {Schmidt}},\ }\href {\doibase 10.1103/PhysRevLett.134.076903} {\bibfield  {journal} {\bibinfo  {journal} {Phys. Rev. Lett.}\ }\textbf {\bibinfo {volume} {134}},\ \bibinfo {pages} {076903} (\bibinfo {year} {2025})}\BibitemShut {NoStop}%
\bibitem [{\citenamefont {Byrnes}\ \emph {et~al.}(2014)\citenamefont {Byrnes}, \citenamefont {Kim},\ and\ \citenamefont {Yamamoto}}]{Yoshihisa2014polariton}%
  \BibitemOpen
  \bibfield  {author} {\bibinfo {author} {\bibfnamefont {T.}~\bibnamefont {Byrnes}}, \bibinfo {author} {\bibfnamefont {N.~Y.}\ \bibnamefont {Kim}}, \ and\ \bibinfo {author} {\bibfnamefont {Y.}~\bibnamefont {Yamamoto}},\ }\href {\doibase https://doi.org/10.1038/nphys3143} {\bibfield  {journal} {\bibinfo  {journal} {Nature Physics}\ }\textbf {\bibinfo {volume} {10}},\ \bibinfo {pages} {803} (\bibinfo {year} {2014})}\BibitemShut {NoStop}%
\bibitem [{\citenamefont {Keeling}\ and\ \citenamefont {K{\'e}na-Cohen}(2020)}]{keeling2020polariton}%
  \BibitemOpen
  \bibfield  {author} {\bibinfo {author} {\bibfnamefont {J.}~\bibnamefont {Keeling}}\ and\ \bibinfo {author} {\bibfnamefont {S.}~\bibnamefont {K{\'e}na-Cohen}},\ }\href {\doibase https://doi.org/10.1146/annurev-physchem-010920-102509} {\bibfield  {journal} {\bibinfo  {journal} {Annual Review of Physical Chemistry}\ }\textbf {\bibinfo {volume} {71}},\ \bibinfo {pages} {435} (\bibinfo {year} {2020})}\BibitemShut {NoStop}%
\bibitem [{\citenamefont {Ghosh}\ \emph {et~al.}(2022)\citenamefont {Ghosh}, \citenamefont {Su}, \citenamefont {Zhao}, \citenamefont {Fieramosca}, \citenamefont {Wu}, \citenamefont {Li}, \citenamefont {Zhang}, \citenamefont {Li}, \citenamefont {Chen}, \citenamefont {Liew} \emph {et~al.}}]{ghosh2022polariton}%
  \BibitemOpen
  \bibfield  {author} {\bibinfo {author} {\bibfnamefont {S.}~\bibnamefont {Ghosh}}, \bibinfo {author} {\bibfnamefont {R.}~\bibnamefont {Su}}, \bibinfo {author} {\bibfnamefont {J.}~\bibnamefont {Zhao}}, \bibinfo {author} {\bibfnamefont {A.}~\bibnamefont {Fieramosca}}, \bibinfo {author} {\bibfnamefont {J.}~\bibnamefont {Wu}}, \bibinfo {author} {\bibfnamefont {T.}~\bibnamefont {Li}}, \bibinfo {author} {\bibfnamefont {Q.}~\bibnamefont {Zhang}}, \bibinfo {author} {\bibfnamefont {F.}~\bibnamefont {Li}}, \bibinfo {author} {\bibfnamefont {Z.}~\bibnamefont {Chen}}, \bibinfo {author} {\bibfnamefont {T.}~\bibnamefont {Liew}},  \emph {et~al.},\ }\href {\doibase https://doi.org/10.3788/PI.2022.R04} {\bibfield  {journal} {\bibinfo  {journal} {Photonics Insights}\ }\textbf {\bibinfo {volume} {1}},\ \bibinfo {pages} {R04} (\bibinfo {year} {2022})}\BibitemShut {NoStop}%
\bibitem [{\citenamefont {Bloch}\ \emph {et~al.}(2008)\citenamefont {Bloch}, \citenamefont {Dalibard},\ and\ \citenamefont {Zwerger}}]{RevModPhys.80.885}%
  \BibitemOpen
  \bibfield  {author} {\bibinfo {author} {\bibfnamefont {I.}~\bibnamefont {Bloch}}, \bibinfo {author} {\bibfnamefont {J.}~\bibnamefont {Dalibard}}, \ and\ \bibinfo {author} {\bibfnamefont {W.}~\bibnamefont {Zwerger}},\ }\href {\doibase 10.1103/RevModPhys.80.885} {\bibfield  {journal} {\bibinfo  {journal} {Rev. Mod. Phys.}\ }\textbf {\bibinfo {volume} {80}},\ \bibinfo {pages} {885} (\bibinfo {year} {2008})}\BibitemShut {NoStop}%
\bibitem [{\citenamefont {Gross}\ and\ \citenamefont {Bloch}(2017)}]{doi:10.1126/science.aal3837}%
  \BibitemOpen
  \bibfield  {author} {\bibinfo {author} {\bibfnamefont {C.}~\bibnamefont {Gross}}\ and\ \bibinfo {author} {\bibfnamefont {I.}~\bibnamefont {Bloch}},\ }\href {\doibase 10.1126/science.aal3837} {\bibfield  {journal} {\bibinfo  {journal} {Science}\ }\textbf {\bibinfo {volume} {357}},\ \bibinfo {pages} {995} (\bibinfo {year} {2017})}\BibitemShut {NoStop}%
\bibitem [{\citenamefont {Schfer}\ \emph {et~al.}(2020)\citenamefont {Schfer}, \citenamefont {Fukuhara}, \citenamefont {Sugawa}, \citenamefont {Takasu},\ and\ \citenamefont {Takahashi}}]{Tools}%
  \BibitemOpen
  \bibfield  {author} {\bibinfo {author} {\bibfnamefont {F.}~\bibnamefont {Schfer}}, \bibinfo {author} {\bibfnamefont {T.}~\bibnamefont {Fukuhara}}, \bibinfo {author} {\bibfnamefont {S.}~\bibnamefont {Sugawa}}, \bibinfo {author} {\bibfnamefont {Y.}~\bibnamefont {Takasu}}, \ and\ \bibinfo {author} {\bibfnamefont {Y.}~\bibnamefont {Takahashi}},\ }\href {https://doi.org/10.1038/s42254-020-0195-3} {\bibfield  {journal} {\bibinfo  {journal} {Nature Reviews Physics}\ }\textbf {\bibinfo {volume} {2}},\ \bibinfo {pages} {411} (\bibinfo {year} {2020})}\BibitemShut {NoStop}%
\bibitem [{\citenamefont {Scazza}\ \emph {et~al.}(2022)\citenamefont {Scazza}, \citenamefont {Zaccanti}, \citenamefont {Massignan}, \citenamefont {Parish},\ and\ \citenamefont {Levinsen}}]{atoms10020055polaron}%
  \BibitemOpen
  \bibfield  {author} {\bibinfo {author} {\bibfnamefont {F.}~\bibnamefont {Scazza}}, \bibinfo {author} {\bibfnamefont {M.}~\bibnamefont {Zaccanti}}, \bibinfo {author} {\bibfnamefont {P.}~\bibnamefont {Massignan}}, \bibinfo {author} {\bibfnamefont {M.~M.}\ \bibnamefont {Parish}}, \ and\ \bibinfo {author} {\bibfnamefont {J.}~\bibnamefont {Levinsen}},\ }\href {https://www.mdpi.com/2218-2004/10/2/55} {\bibfield  {journal} {\bibinfo  {journal} {Atoms}\ }\textbf {\bibinfo {volume} {10}},\ \bibinfo {pages} {55} (\bibinfo {year} {2022})}\BibitemShut {NoStop}%
\bibitem [{\citenamefont {Baroni}\ \emph {et~al.}(2024)\citenamefont {Baroni}, \citenamefont {Huang}, \citenamefont {Fritsche}, \citenamefont {Dobler}, \citenamefont {Anich}, \citenamefont {Kirilov}, \citenamefont {Grimm}, \citenamefont {Bastarrachea-Magnani}, \citenamefont {Massignan},\ and\ \citenamefont {Bruun}}]{baroni2024polaron}%
  \BibitemOpen
  \bibfield  {author} {\bibinfo {author} {\bibfnamefont {C.}~\bibnamefont {Baroni}}, \bibinfo {author} {\bibfnamefont {B.}~\bibnamefont {Huang}}, \bibinfo {author} {\bibfnamefont {I.}~\bibnamefont {Fritsche}}, \bibinfo {author} {\bibfnamefont {E.}~\bibnamefont {Dobler}}, \bibinfo {author} {\bibfnamefont {G.}~\bibnamefont {Anich}}, \bibinfo {author} {\bibfnamefont {E.}~\bibnamefont {Kirilov}}, \bibinfo {author} {\bibfnamefont {R.}~\bibnamefont {Grimm}}, \bibinfo {author} {\bibfnamefont {M.~A.}\ \bibnamefont {Bastarrachea-Magnani}}, \bibinfo {author} {\bibfnamefont {P.}~\bibnamefont {Massignan}}, \ and\ \bibinfo {author} {\bibfnamefont {G.~M.}\ \bibnamefont {Bruun}},\ }\href {\doibase https://doi.org/10.1038/s41567-023-02248-4} {\bibfield  {journal} {\bibinfo  {journal} {Nature Physics}\ }\textbf {\bibinfo {volume} {20}},\ \bibinfo {pages} {68} (\bibinfo {year} {2024})}\BibitemShut {NoStop}%
\bibitem [{\citenamefont {Fritsche}\ \emph {et~al.}(2021)\citenamefont {Fritsche}, \citenamefont {Baroni}, \citenamefont {Dobler}, \citenamefont {Kirilov}, \citenamefont {Huang}, \citenamefont {Grimm}, \citenamefont {Bruun},\ and\ \citenamefont {Massignan}}]{BFMpolaron2021exp}%
  \BibitemOpen
  \bibfield  {author} {\bibinfo {author} {\bibfnamefont {I.}~\bibnamefont {Fritsche}}, \bibinfo {author} {\bibfnamefont {C.}~\bibnamefont {Baroni}}, \bibinfo {author} {\bibfnamefont {E.}~\bibnamefont {Dobler}}, \bibinfo {author} {\bibfnamefont {E.}~\bibnamefont {Kirilov}}, \bibinfo {author} {\bibfnamefont {B.}~\bibnamefont {Huang}}, \bibinfo {author} {\bibfnamefont {R.}~\bibnamefont {Grimm}}, \bibinfo {author} {\bibfnamefont {G.~M.}\ \bibnamefont {Bruun}}, \ and\ \bibinfo {author} {\bibfnamefont {P.}~\bibnamefont {Massignan}},\ }\href {\doibase 10.1103/PhysRevA.103.053314} {\bibfield  {journal} {\bibinfo  {journal} {Phys. Rev. A}\ }\textbf {\bibinfo {volume} {103}},\ \bibinfo {pages} {053314} (\bibinfo {year} {2021})}\BibitemShut {NoStop}%
\bibitem [{\citenamefont {Stan}\ \emph {et~al.}(2004)\citenamefont {Stan}, \citenamefont {Zwierlein}, \citenamefont {Schunck}, \citenamefont {Raupach},\ and\ \citenamefont {Ketterle}}]{PhysRevLett.93.143001}%
  \BibitemOpen
  \bibfield  {author} {\bibinfo {author} {\bibfnamefont {C.~A.}\ \bibnamefont {Stan}}, \bibinfo {author} {\bibfnamefont {M.~W.}\ \bibnamefont {Zwierlein}}, \bibinfo {author} {\bibfnamefont {C.~H.}\ \bibnamefont {Schunck}}, \bibinfo {author} {\bibfnamefont {S.~M.~F.}\ \bibnamefont {Raupach}}, \ and\ \bibinfo {author} {\bibfnamefont {W.}~\bibnamefont {Ketterle}},\ }\href {\doibase 10.1103/PhysRevLett.93.143001} {\bibfield  {journal} {\bibinfo  {journal} {Phys. Rev. Lett.}\ }\textbf {\bibinfo {volume} {93}},\ \bibinfo {pages} {143001} (\bibinfo {year} {2004})}\BibitemShut {NoStop}%
\bibitem [{\citenamefont {Milczewski}\ and\ \citenamefont {Duda}(2023)}]{milczewski2023molecules}%
  \BibitemOpen
  \bibfield  {author} {\bibinfo {author} {\bibfnamefont {J.~v.}\ \bibnamefont {Milczewski}}\ and\ \bibinfo {author} {\bibfnamefont {M.}~\bibnamefont {Duda}},\ }\href {\doibase https://doi.org/10.1038/s41567-023-01950-7} {\bibfield  {journal} {\bibinfo  {journal} {Nature Physics}\ }\textbf {\bibinfo {volume} {19}},\ \bibinfo {pages} {624} (\bibinfo {year} {2023})}\BibitemShut {NoStop}%
\bibitem [{\citenamefont {Duda}\ \emph {et~al.}(2023)\citenamefont {Duda}, \citenamefont {Chen}, \citenamefont {Schindewolf}, \citenamefont {Bause}, \citenamefont {von Milczewski}, \citenamefont {Schmidt}, \citenamefont {Bloch},\ and\ \citenamefont {Luo}}]{duda2023molecules}%
  \BibitemOpen
  \bibfield  {author} {\bibinfo {author} {\bibfnamefont {M.}~\bibnamefont {Duda}}, \bibinfo {author} {\bibfnamefont {X.-Y.}\ \bibnamefont {Chen}}, \bibinfo {author} {\bibfnamefont {A.}~\bibnamefont {Schindewolf}}, \bibinfo {author} {\bibfnamefont {R.}~\bibnamefont {Bause}}, \bibinfo {author} {\bibfnamefont {J.}~\bibnamefont {von Milczewski}}, \bibinfo {author} {\bibfnamefont {R.}~\bibnamefont {Schmidt}}, \bibinfo {author} {\bibfnamefont {I.}~\bibnamefont {Bloch}}, \ and\ \bibinfo {author} {\bibfnamefont {X.-Y.}\ \bibnamefont {Luo}},\ }\href {\doibase https://doi.org/10.1038/s41567-023-01948-1} {\bibfield  {journal} {\bibinfo  {journal} {Nature Physics}\ }\textbf {\bibinfo {volume} {19}},\ \bibinfo {pages} {720} (\bibinfo {year} {2023})}\BibitemShut {NoStop}%
\bibitem [{\citenamefont {Ferrier-Barbut}\ \emph {et~al.}(2014)\citenamefont {Ferrier-Barbut}, \citenamefont {Delehaye}, \citenamefont {Laurent}, \citenamefont {Grier}, \citenamefont {Pierce}, \citenamefont {Rem}, \citenamefont {Chevy},\ and\ \citenamefont {Salomon}}]{doi:10.1126/science.1255380}%
  \BibitemOpen
  \bibfield  {author} {\bibinfo {author} {\bibfnamefont {I.}~\bibnamefont {Ferrier-Barbut}}, \bibinfo {author} {\bibfnamefont {M.}~\bibnamefont {Delehaye}}, \bibinfo {author} {\bibfnamefont {S.}~\bibnamefont {Laurent}}, \bibinfo {author} {\bibfnamefont {A.~T.}\ \bibnamefont {Grier}}, \bibinfo {author} {\bibfnamefont {M.}~\bibnamefont {Pierce}}, \bibinfo {author} {\bibfnamefont {B.~S.}\ \bibnamefont {Rem}}, \bibinfo {author} {\bibfnamefont {F.}~\bibnamefont {Chevy}}, \ and\ \bibinfo {author} {\bibfnamefont {C.}~\bibnamefont {Salomon}},\ }\href {\doibase 10.1126/science.1255380} {\bibfield  {journal} {\bibinfo  {journal} {Science}\ }\textbf {\bibinfo {volume} {345}},\ \bibinfo {pages} {1035} (\bibinfo {year} {2014})}\BibitemShut {NoStop}%
\bibitem [{\citenamefont {Roy}\ \emph {et~al.}(2017)\citenamefont {Roy}, \citenamefont {Green}, \citenamefont {Bowler},\ and\ \citenamefont {Gupta}}]{PhysRevLett.118.055301}%
  \BibitemOpen
  \bibfield  {author} {\bibinfo {author} {\bibfnamefont {R.}~\bibnamefont {Roy}}, \bibinfo {author} {\bibfnamefont {A.}~\bibnamefont {Green}}, \bibinfo {author} {\bibfnamefont {R.}~\bibnamefont {Bowler}}, \ and\ \bibinfo {author} {\bibfnamefont {S.}~\bibnamefont {Gupta}},\ }\href {\doibase 10.1103/PhysRevLett.118.055301} {\bibfield  {journal} {\bibinfo  {journal} {Phys. Rev. Lett.}\ }\textbf {\bibinfo {volume} {118}},\ \bibinfo {pages} {055301} (\bibinfo {year} {2017})}\BibitemShut {NoStop}%
\bibitem [{\citenamefont {Yao}\ \emph {et~al.}(2016)\citenamefont {Yao}, \citenamefont {Chen}, \citenamefont {Wu}, \citenamefont {Liu}, \citenamefont {Wang}, \citenamefont {Jiang}, \citenamefont {Deng}, \citenamefont {Chen},\ and\ \citenamefont {Pan}}]{doubleSF2016}%
  \BibitemOpen
  \bibfield  {author} {\bibinfo {author} {\bibfnamefont {X.-C.}\ \bibnamefont {Yao}}, \bibinfo {author} {\bibfnamefont {H.-Z.}\ \bibnamefont {Chen}}, \bibinfo {author} {\bibfnamefont {Y.-P.}\ \bibnamefont {Wu}}, \bibinfo {author} {\bibfnamefont {X.-P.}\ \bibnamefont {Liu}}, \bibinfo {author} {\bibfnamefont {X.-Q.}\ \bibnamefont {Wang}}, \bibinfo {author} {\bibfnamefont {X.}~\bibnamefont {Jiang}}, \bibinfo {author} {\bibfnamefont {Y.}~\bibnamefont {Deng}}, \bibinfo {author} {\bibfnamefont {Y.-A.}\ \bibnamefont {Chen}}, \ and\ \bibinfo {author} {\bibfnamefont {J.-W.}\ \bibnamefont {Pan}},\ }\href {\doibase 10.1103/PhysRevLett.117.145301} {\bibfield  {journal} {\bibinfo  {journal} {Phys. Rev. Lett.}\ }\textbf {\bibinfo {volume} {117}},\ \bibinfo {pages} {145301} (\bibinfo {year} {2016})}\BibitemShut {NoStop}%
\bibitem [{\citenamefont {Pollet}\ \emph {et~al.}(2006)\citenamefont {Pollet}, \citenamefont {Troyer}, \citenamefont {Van~Houcke},\ and\ \citenamefont {Rombouts}}]{bfmPhaseDiagram2006}%
  \BibitemOpen
  \bibfield  {author} {\bibinfo {author} {\bibfnamefont {L.}~\bibnamefont {Pollet}}, \bibinfo {author} {\bibfnamefont {M.}~\bibnamefont {Troyer}}, \bibinfo {author} {\bibfnamefont {K.}~\bibnamefont {Van~Houcke}}, \ and\ \bibinfo {author} {\bibfnamefont {S.~M.~A.}\ \bibnamefont {Rombouts}},\ }\href {\doibase 10.1103/PhysRevLett.96.190402} {\bibfield  {journal} {\bibinfo  {journal} {Phys. Rev. Lett.}\ }\textbf {\bibinfo {volume} {96}},\ \bibinfo {pages} {190402} (\bibinfo {year} {2006})}\BibitemShut {NoStop}%
\bibitem [{\citenamefont {Cazalilla}\ and\ \citenamefont {Ho}(2003)}]{PhysRevLett.91.150403}%
  \BibitemOpen
  \bibfield  {author} {\bibinfo {author} {\bibfnamefont {M.~A.}\ \bibnamefont {Cazalilla}}\ and\ \bibinfo {author} {\bibfnamefont {A.~F.}\ \bibnamefont {Ho}},\ }\href {\doibase 10.1103/PhysRevLett.91.150403} {\bibfield  {journal} {\bibinfo  {journal} {Phys. Rev. Lett.}\ }\textbf {\bibinfo {volume} {91}},\ \bibinfo {pages} {150403} (\bibinfo {year} {2003})}\BibitemShut {NoStop}%
\bibitem [{\citenamefont {Mathey}\ \emph {et~al.}(2004)\citenamefont {Mathey}, \citenamefont {Wang}, \citenamefont {Hofstetter}, \citenamefont {Lukin},\ and\ \citenamefont {Demler}}]{PhysRevLett.93.120404}%
  \BibitemOpen
  \bibfield  {author} {\bibinfo {author} {\bibfnamefont {L.}~\bibnamefont {Mathey}}, \bibinfo {author} {\bibfnamefont {D.-W.}\ \bibnamefont {Wang}}, \bibinfo {author} {\bibfnamefont {W.}~\bibnamefont {Hofstetter}}, \bibinfo {author} {\bibfnamefont {M.~D.}\ \bibnamefont {Lukin}}, \ and\ \bibinfo {author} {\bibfnamefont {E.}~\bibnamefont {Demler}},\ }\href {\doibase 10.1103/PhysRevLett.93.120404} {\bibfield  {journal} {\bibinfo  {journal} {Phys. Rev. Lett.}\ }\textbf {\bibinfo {volume} {93}},\ \bibinfo {pages} {120404} (\bibinfo {year} {2004})}\BibitemShut {NoStop}%
\bibitem [{\citenamefont {Barillier-Pertuisel}\ \emph {et~al.}(2008)\citenamefont {Barillier-Pertuisel}, \citenamefont {Pittel}, \citenamefont {Pollet},\ and\ \citenamefont {Schuck}}]{PhysRevA.77.012115}%
  \BibitemOpen
  \bibfield  {author} {\bibinfo {author} {\bibfnamefont {X.}~\bibnamefont {Barillier-Pertuisel}}, \bibinfo {author} {\bibfnamefont {S.}~\bibnamefont {Pittel}}, \bibinfo {author} {\bibfnamefont {L.}~\bibnamefont {Pollet}}, \ and\ \bibinfo {author} {\bibfnamefont {P.}~\bibnamefont {Schuck}},\ }\href {\doibase 10.1103/PhysRevA.77.012115} {\bibfield  {journal} {\bibinfo  {journal} {Phys. Rev. A}\ }\textbf {\bibinfo {volume} {77}},\ \bibinfo {pages} {012115} (\bibinfo {year} {2008})}\BibitemShut {NoStop}%
\bibitem [{\citenamefont {Mering}\ and\ \citenamefont {Fleischhauer}(2008)}]{PhysRevA.77.023601}%
  \BibitemOpen
  \bibfield  {author} {\bibinfo {author} {\bibfnamefont {A.}~\bibnamefont {Mering}}\ and\ \bibinfo {author} {\bibfnamefont {M.}~\bibnamefont {Fleischhauer}},\ }\href {\doibase 10.1103/PhysRevA.77.023601} {\bibfield  {journal} {\bibinfo  {journal} {Phys. Rev. A}\ }\textbf {\bibinfo {volume} {77}},\ \bibinfo {pages} {023601} (\bibinfo {year} {2008})}\BibitemShut {NoStop}%
\bibitem [{\citenamefont {Pollet}\ \emph {et~al.}(2008)\citenamefont {Pollet}, \citenamefont {Kollath}, \citenamefont {Schollw\"ock},\ and\ \citenamefont {Troyer}}]{PhysRevA.77.023608}%
  \BibitemOpen
  \bibfield  {author} {\bibinfo {author} {\bibfnamefont {L.}~\bibnamefont {Pollet}}, \bibinfo {author} {\bibfnamefont {C.}~\bibnamefont {Kollath}}, \bibinfo {author} {\bibfnamefont {U.}~\bibnamefont {Schollw\"ock}}, \ and\ \bibinfo {author} {\bibfnamefont {M.}~\bibnamefont {Troyer}},\ }\href {\doibase 10.1103/PhysRevA.77.023608} {\bibfield  {journal} {\bibinfo  {journal} {Phys. Rev. A}\ }\textbf {\bibinfo {volume} {77}},\ \bibinfo {pages} {023608} (\bibinfo {year} {2008})}\BibitemShut {NoStop}%
\bibitem [{\citenamefont {Kagan}\ \emph {et~al.}(2004)\citenamefont {Kagan}, \citenamefont {Brodsky}, \citenamefont {Efremov},\ and\ \citenamefont {Klaptsov}}]{trio2004}%
  \BibitemOpen
  \bibfield  {author} {\bibinfo {author} {\bibfnamefont {M.~Y.}\ \bibnamefont {Kagan}}, \bibinfo {author} {\bibfnamefont {I.~V.}\ \bibnamefont {Brodsky}}, \bibinfo {author} {\bibfnamefont {D.~V.}\ \bibnamefont {Efremov}}, \ and\ \bibinfo {author} {\bibfnamefont {A.~V.}\ \bibnamefont {Klaptsov}},\ }\href {\doibase 10.1103/PhysRevA.70.023607} {\bibfield  {journal} {\bibinfo  {journal} {Phys. Rev. A}\ }\textbf {\bibinfo {volume} {70}},\ \bibinfo {pages} {023607} (\bibinfo {year} {2004})}\BibitemShut {NoStop}%
\bibitem [{\citenamefont {Yang}\ \emph {et~al.}(2022{\natexlab{a}})\citenamefont {Yang}, \citenamefont {Cao}, \citenamefont {Su}, \citenamefont {Rui}, \citenamefont {Zhao},\ and\ \citenamefont {Pan}}]{Pan2022CreationAtomMoleculeNaKK}%
  \BibitemOpen
  \bibfield  {author} {\bibinfo {author} {\bibfnamefont {H.}~\bibnamefont {Yang}}, \bibinfo {author} {\bibfnamefont {J.}~\bibnamefont {Cao}}, \bibinfo {author} {\bibfnamefont {Z.}~\bibnamefont {Su}}, \bibinfo {author} {\bibfnamefont {J.}~\bibnamefont {Rui}}, \bibinfo {author} {\bibfnamefont {B.}~\bibnamefont {Zhao}}, \ and\ \bibinfo {author} {\bibfnamefont {J.-W.}\ \bibnamefont {Pan}},\ }\href {\doibase https://doi.org/10.1126/science.ade6307} {\bibfield  {journal} {\bibinfo  {journal} {Science}\ }\textbf {\bibinfo {volume} {378}},\ \bibinfo {pages} {1009} (\bibinfo {year} {2022}{\natexlab{a}})}\BibitemShut {NoStop}%
\bibitem [{\citenamefont {Yang}\ \emph {et~al.}(2022{\natexlab{b}})\citenamefont {Yang}, \citenamefont {Wang}, \citenamefont {Su}, \citenamefont {Cao}, \citenamefont {Zhang}, \citenamefont {Rui}, \citenamefont {Zhao}, \citenamefont {Bai},\ and\ \citenamefont {Pan}}]{Pan2022EvidenceAtomMoleculeNaKK}%
  \BibitemOpen
  \bibfield  {author} {\bibinfo {author} {\bibfnamefont {H.}~\bibnamefont {Yang}}, \bibinfo {author} {\bibfnamefont {X.-Y.}\ \bibnamefont {Wang}}, \bibinfo {author} {\bibfnamefont {Z.}~\bibnamefont {Su}}, \bibinfo {author} {\bibfnamefont {J.}~\bibnamefont {Cao}}, \bibinfo {author} {\bibfnamefont {D.-C.}\ \bibnamefont {Zhang}}, \bibinfo {author} {\bibfnamefont {J.}~\bibnamefont {Rui}}, \bibinfo {author} {\bibfnamefont {B.}~\bibnamefont {Zhao}}, \bibinfo {author} {\bibfnamefont {C.-L.}\ \bibnamefont {Bai}}, \ and\ \bibinfo {author} {\bibfnamefont {J.-W.}\ \bibnamefont {Pan}},\ }\href {\doibase https://doi.org/10.1038/s41586-021-04297-2} {\bibfield  {journal} {\bibinfo  {journal} {Nature}\ }\textbf {\bibinfo {volume} {602}},\ \bibinfo {pages} {229} (\bibinfo {year} {2022}{\natexlab{b}})}\BibitemShut {NoStop}%
\bibitem [{\citenamefont {Cao}\ \emph {et~al.}(2024)\citenamefont {Cao}, \citenamefont {Wang}, \citenamefont {Yang}, \citenamefont {Fan}, \citenamefont {Su}, \citenamefont {Rui}, \citenamefont {Zhao},\ and\ \citenamefont {Pan}}]{Pan2024PhotoassociationNaKK}%
  \BibitemOpen
  \bibfield  {author} {\bibinfo {author} {\bibfnamefont {J.}~\bibnamefont {Cao}}, \bibinfo {author} {\bibfnamefont {B.-Y.}\ \bibnamefont {Wang}}, \bibinfo {author} {\bibfnamefont {H.}~\bibnamefont {Yang}}, \bibinfo {author} {\bibfnamefont {Z.-J.}\ \bibnamefont {Fan}}, \bibinfo {author} {\bibfnamefont {Z.}~\bibnamefont {Su}}, \bibinfo {author} {\bibfnamefont {J.}~\bibnamefont {Rui}}, \bibinfo {author} {\bibfnamefont {B.}~\bibnamefont {Zhao}}, \ and\ \bibinfo {author} {\bibfnamefont {J.-W.}\ \bibnamefont {Pan}},\ }\href {\doibase 10.1103/PhysRevLett.132.093403} {\bibfield  {journal} {\bibinfo  {journal} {Phys. Rev. Lett.}\ }\textbf {\bibinfo {volume} {132}},\ \bibinfo {pages} {093403} (\bibinfo {year} {2024})}\BibitemShut {NoStop}%
\bibitem [{\citenamefont {Ruhman}\ and\ \citenamefont {Altman}(2017)}]{cc32spinless2017}%
  \BibitemOpen
  \bibfield  {author} {\bibinfo {author} {\bibfnamefont {J.}~\bibnamefont {Ruhman}}\ and\ \bibinfo {author} {\bibfnamefont {E.}~\bibnamefont {Altman}},\ }\href {\doibase 10.1103/PhysRevB.96.085133} {\bibfield  {journal} {\bibinfo  {journal} {Phys. Rev. B}\ }\textbf {\bibinfo {volume} {96}},\ \bibinfo {pages} {085133} (\bibinfo {year} {2017})}\BibitemShut {NoStop}%
\bibitem [{\citenamefont {Gotta}\ \emph {et~al.}(2021{\natexlab{a}})\citenamefont {Gotta}, \citenamefont {Mazza}, \citenamefont {Simon},\ and\ \citenamefont {Roux}}]{TwoFluid2021}%
  \BibitemOpen
  \bibfield  {author} {\bibinfo {author} {\bibfnamefont {L.}~\bibnamefont {Gotta}}, \bibinfo {author} {\bibfnamefont {L.}~\bibnamefont {Mazza}}, \bibinfo {author} {\bibfnamefont {P.}~\bibnamefont {Simon}}, \ and\ \bibinfo {author} {\bibfnamefont {G.}~\bibnamefont {Roux}},\ }\href {\doibase 10.1103/PhysRevLett.126.206805} {\bibfield  {journal} {\bibinfo  {journal} {Phys. Rev. Lett.}\ }\textbf {\bibinfo {volume} {126}},\ \bibinfo {pages} {206805} (\bibinfo {year} {2021}{\natexlab{a}})}\BibitemShut {NoStop}%
\bibitem [{\citenamefont {Gotta}\ \emph {et~al.}(2022{\natexlab{a}})\citenamefont {Gotta}, \citenamefont {Mazza}, \citenamefont {Simon},\ and\ \citenamefont {Roux}}]{KineticFpair2022}%
  \BibitemOpen
  \bibfield  {author} {\bibinfo {author} {\bibfnamefont {L.}~\bibnamefont {Gotta}}, \bibinfo {author} {\bibfnamefont {L.}~\bibnamefont {Mazza}}, \bibinfo {author} {\bibfnamefont {P.}~\bibnamefont {Simon}}, \ and\ \bibinfo {author} {\bibfnamefont {G.}~\bibnamefont {Roux}},\ }\href {\doibase 10.1103/PhysRevB.105.134512} {\bibfield  {journal} {\bibinfo  {journal} {Phys. Rev. B}\ }\textbf {\bibinfo {volume} {105}},\ \bibinfo {pages} {134512} (\bibinfo {year} {2022}{\natexlab{a}})}\BibitemShut {NoStop}%
\bibitem [{\citenamefont {Baranov}\ \emph {et~al.}(2012)\citenamefont {Baranov}, \citenamefont {Dalmonte}, \citenamefont {Pupillo},\ and\ \citenamefont {Zoller}}]{Zoller2012dipolarReview}%
  \BibitemOpen
  \bibfield  {author} {\bibinfo {author} {\bibfnamefont {M.~A.}\ \bibnamefont {Baranov}}, \bibinfo {author} {\bibfnamefont {M.}~\bibnamefont {Dalmonte}}, \bibinfo {author} {\bibfnamefont {G.}~\bibnamefont {Pupillo}}, \ and\ \bibinfo {author} {\bibfnamefont {P.}~\bibnamefont {Zoller}},\ }\href {https://pubs.acs.org/doi/10.1021/cr2003568} {\bibfield  {journal} {\bibinfo  {journal} {Chemical Reviews}\ }\textbf {\bibinfo {volume} {112}},\ \bibinfo {pages} {5012} (\bibinfo {year} {2012})}\BibitemShut {NoStop}%
\bibitem [{\citenamefont {Wang}\ \emph {et~al.}(2010)\citenamefont {Wang}, \citenamefont {Wang},\ and\ \citenamefont {Das~Sarma}}]{EBFH2010Wang}%
  \BibitemOpen
  \bibfield  {author} {\bibinfo {author} {\bibfnamefont {B.}~\bibnamefont {Wang}}, \bibinfo {author} {\bibfnamefont {D.-W.}\ \bibnamefont {Wang}}, \ and\ \bibinfo {author} {\bibfnamefont {S.}~\bibnamefont {Das~Sarma}},\ }\href {\doibase 10.1103/PhysRevA.82.021602} {\bibfield  {journal} {\bibinfo  {journal} {Phys. Rev. A}\ }\textbf {\bibinfo {volume} {82}},\ \bibinfo {pages} {021602} (\bibinfo {year} {2010})}\BibitemShut {NoStop}%
\bibitem [{\citenamefont {Orignac}\ \emph {et~al.}(2010)\citenamefont {Orignac}, \citenamefont {Tsuchiizu},\ and\ \citenamefont {Suzumura}}]{PhysRevA.81.053626}%
  \BibitemOpen
  \bibfield  {author} {\bibinfo {author} {\bibfnamefont {E.}~\bibnamefont {Orignac}}, \bibinfo {author} {\bibfnamefont {M.}~\bibnamefont {Tsuchiizu}}, \ and\ \bibinfo {author} {\bibfnamefont {Y.}~\bibnamefont {Suzumura}},\ }\href {\doibase 10.1103/PhysRevA.81.053626} {\bibfield  {journal} {\bibinfo  {journal} {Phys. Rev. A}\ }\textbf {\bibinfo {volume} {81}},\ \bibinfo {pages} {053626} (\bibinfo {year} {2010})}\BibitemShut {NoStop}%
\bibitem [{\citenamefont {Gómez-Lozada}\ \emph {et~al.}(2025)\citenamefont {Gómez-Lozada}, \citenamefont {Franco},\ and\ \citenamefont {Silva-Valencia}}]{EBFH2025SciPost}%
  \BibitemOpen
  \bibfield  {author} {\bibinfo {author} {\bibfnamefont {F.}~\bibnamefont {Gómez-Lozada}}, \bibinfo {author} {\bibfnamefont {R.}~\bibnamefont {Franco}}, \ and\ \bibinfo {author} {\bibfnamefont {J.}~\bibnamefont {Silva-Valencia}},\ }\href {\doibase 10.21468/SciPostPhysCore.8.1.007} {\bibfield  {journal} {\bibinfo  {journal} {SciPost Phys. Core}\ }\textbf {\bibinfo {volume} {8}},\ \bibinfo {pages} {007} (\bibinfo {year} {2025})}\BibitemShut {NoStop}%
\bibitem [{\citenamefont {Carr}\ \emph {et~al.}(2009)\citenamefont {Carr}, \citenamefont {DeMille}, \citenamefont {Krems},\ and\ \citenamefont {Ye}}]{carr2009coldApp}%
  \BibitemOpen
  \bibfield  {author} {\bibinfo {author} {\bibfnamefont {L.~D.}\ \bibnamefont {Carr}}, \bibinfo {author} {\bibfnamefont {D.}~\bibnamefont {DeMille}}, \bibinfo {author} {\bibfnamefont {R.~V.}\ \bibnamefont {Krems}}, \ and\ \bibinfo {author} {\bibfnamefont {J.}~\bibnamefont {Ye}},\ }\href {\doibase 10.1088/1367-2630/11/5/055009} {\bibfield  {journal} {\bibinfo  {journal} {New Journal of Physics}\ }\textbf {\bibinfo {volume} {11}},\ \bibinfo {pages} {055049} (\bibinfo {year} {2009})}\BibitemShut {NoStop}%
\bibitem [{\citenamefont {Bohn}\ \emph {et~al.}(2017)\citenamefont {Bohn}, \citenamefont {Rey},\ and\ \citenamefont {Ye}}]{bohn2017Chemistry}%
  \BibitemOpen
  \bibfield  {author} {\bibinfo {author} {\bibfnamefont {J.~L.}\ \bibnamefont {Bohn}}, \bibinfo {author} {\bibfnamefont {A.~M.}\ \bibnamefont {Rey}}, \ and\ \bibinfo {author} {\bibfnamefont {J.}~\bibnamefont {Ye}},\ }\href {\doibase https://doi.org/10.1126/science.aam6299} {\bibfield  {journal} {\bibinfo  {journal} {Science}\ }\textbf {\bibinfo {volume} {357}},\ \bibinfo {pages} {1002} (\bibinfo {year} {2017})}\BibitemShut {NoStop}%
\bibitem [{\citenamefont {Gotta}\ \emph {et~al.}(2021{\natexlab{b}})\citenamefont {Gotta}, \citenamefont {Mazza}, \citenamefont {Simon},\ and\ \citenamefont {Roux}}]{pairHopping2021spinless}%
  \BibitemOpen
  \bibfield  {author} {\bibinfo {author} {\bibfnamefont {L.}~\bibnamefont {Gotta}}, \bibinfo {author} {\bibfnamefont {L.}~\bibnamefont {Mazza}}, \bibinfo {author} {\bibfnamefont {P.}~\bibnamefont {Simon}}, \ and\ \bibinfo {author} {\bibfnamefont {G.}~\bibnamefont {Roux}},\ }\href {\doibase 10.1103/PhysRevB.104.094521} {\bibfield  {journal} {\bibinfo  {journal} {Phys. Rev. B}\ }\textbf {\bibinfo {volume} {104}},\ \bibinfo {pages} {094521} (\bibinfo {year} {2021}{\natexlab{b}})}\BibitemShut {NoStop}%
\bibitem [{\citenamefont {Gotta}\ \emph {et~al.}(2022{\natexlab{b}})\citenamefont {Gotta}, \citenamefont {Mazza}, \citenamefont {Simon},\ and\ \citenamefont {Roux}}]{pairHopping2022spinless}%
  \BibitemOpen
  \bibfield  {author} {\bibinfo {author} {\bibfnamefont {L.}~\bibnamefont {Gotta}}, \bibinfo {author} {\bibfnamefont {L.}~\bibnamefont {Mazza}}, \bibinfo {author} {\bibfnamefont {P.}~\bibnamefont {Simon}}, \ and\ \bibinfo {author} {\bibfnamefont {G.}~\bibnamefont {Roux}},\ }\href {\doibase 10.1103/PhysRevB.106.235147} {\bibfield  {journal} {\bibinfo  {journal} {Phys. Rev. B}\ }\textbf {\bibinfo {volume} {106}},\ \bibinfo {pages} {235147} (\bibinfo {year} {2022}{\natexlab{b}})}\BibitemShut {NoStop}%
\bibitem [{\citenamefont {Bariev}(1991)}]{bariev1991spinless}%
  \BibitemOpen
  \bibfield  {author} {\bibinfo {author} {\bibfnamefont {R.}~\bibnamefont {Bariev}},\ }\href@noop {} {\bibfield  {journal} {\bibinfo  {journal} {Journal of Physics A: Mathematical and General}\ }\textbf {\bibinfo {volume} {24}},\ \bibinfo {pages} {L549} (\bibinfo {year} {1991})}\BibitemShut {NoStop}%
\bibitem [{\citenamefont {White}(1992)}]{PhysRevLett.69.2863}%
  \BibitemOpen
  \bibfield  {author} {\bibinfo {author} {\bibfnamefont {S.~R.}\ \bibnamefont {White}},\ }\href {\doibase 10.1103/PhysRevLett.69.2863} {\bibfield  {journal} {\bibinfo  {journal} {Phys. Rev. Lett.}\ }\textbf {\bibinfo {volume} {69}},\ \bibinfo {pages} {2863} (\bibinfo {year} {1992})}\BibitemShut {NoStop}%
\bibitem [{\citenamefont {Fishman}\ \emph {et~al.}(2022)\citenamefont {Fishman}, \citenamefont {White},\ and\ \citenamefont {Stoudenmire}}]{ITensor}%
  \BibitemOpen
  \bibfield  {author} {\bibinfo {author} {\bibfnamefont {M.}~\bibnamefont {Fishman}}, \bibinfo {author} {\bibfnamefont {S.~R.}\ \bibnamefont {White}}, \ and\ \bibinfo {author} {\bibfnamefont {E.~M.}\ \bibnamefont {Stoudenmire}},\ }\href {\doibase 10.21468/SciPostPhysCodeb.4} {\bibfield  {journal} {\bibinfo  {journal} {SciPost Phys. Codebases}\ ,\ \bibinfo {pages} {4}} (\bibinfo {year} {2022})}\BibitemShut {NoStop}%
\bibitem [{\citenamefont {Kuklov}\ and\ \citenamefont {Svistunov}(2003)}]{perturbation2003}%
  \BibitemOpen
  \bibfield  {author} {\bibinfo {author} {\bibfnamefont {A.~B.}\ \bibnamefont {Kuklov}}\ and\ \bibinfo {author} {\bibfnamefont {B.~V.}\ \bibnamefont {Svistunov}},\ }\href {\doibase 10.1103/PhysRevLett.90.100401} {\bibfield  {journal} {\bibinfo  {journal} {Phys. Rev. Lett.}\ }\textbf {\bibinfo {volume} {90}},\ \bibinfo {pages} {100401} (\bibinfo {year} {2003})}\BibitemShut {NoStop}%
\bibitem [{\citenamefont {Mondal}\ \emph {et~al.}(2021)\citenamefont {Mondal}, \citenamefont {Greschner}, \citenamefont {Santos},\ and\ \citenamefont {Mishra}}]{perturbationUpSSH2021}%
  \BibitemOpen
  \bibfield  {author} {\bibinfo {author} {\bibfnamefont {S.}~\bibnamefont {Mondal}}, \bibinfo {author} {\bibfnamefont {S.}~\bibnamefont {Greschner}}, \bibinfo {author} {\bibfnamefont {L.}~\bibnamefont {Santos}}, \ and\ \bibinfo {author} {\bibfnamefont {T.}~\bibnamefont {Mishra}},\ }\href {\doibase 10.1103/PhysRevA.104.013315} {\bibfield  {journal} {\bibinfo  {journal} {Phys. Rev. A}\ }\textbf {\bibinfo {volume} {104}},\ \bibinfo {pages} {013315} (\bibinfo {year} {2021})}\BibitemShut {NoStop}%
\bibitem [{\citenamefont {Rizzi}\ and\ \citenamefont {Imambekov}(2008)}]{PhysRevA.77.023621}%
  \BibitemOpen
  \bibfield  {author} {\bibinfo {author} {\bibfnamefont {M.}~\bibnamefont {Rizzi}}\ and\ \bibinfo {author} {\bibfnamefont {A.}~\bibnamefont {Imambekov}},\ }\href {\doibase 10.1103/PhysRevA.77.023621} {\bibfield  {journal} {\bibinfo  {journal} {Phys. Rev. A}\ }\textbf {\bibinfo {volume} {77}},\ \bibinfo {pages} {023621} (\bibinfo {year} {2008})}\BibitemShut {NoStop}%
\bibitem [{\citenamefont {He}\ \emph {et~al.}(2019)\citenamefont {He}, \citenamefont {Tian}, \citenamefont {Pekker},\ and\ \citenamefont {Mong}}]{fpaired2019}%
  \BibitemOpen
  \bibfield  {author} {\bibinfo {author} {\bibfnamefont {Y.}~\bibnamefont {He}}, \bibinfo {author} {\bibfnamefont {B.}~\bibnamefont {Tian}}, \bibinfo {author} {\bibfnamefont {D.}~\bibnamefont {Pekker}}, \ and\ \bibinfo {author} {\bibfnamefont {R.~S.~K.}\ \bibnamefont {Mong}},\ }\href {\doibase 10.1103/PhysRevB.100.201101} {\bibfield  {journal} {\bibinfo  {journal} {Phys. Rev. B}\ }\textbf {\bibinfo {volume} {100}},\ \bibinfo {pages} {201101} (\bibinfo {year} {2019})}\BibitemShut {NoStop}%
\bibitem [{\citenamefont {Arrigoni}\ \emph {et~al.}(2004)\citenamefont {Arrigoni}, \citenamefont {Fradkin},\ and\ \citenamefont {Kivelson}}]{Kivelson2004}%
  \BibitemOpen
  \bibfield  {author} {\bibinfo {author} {\bibfnamefont {E.}~\bibnamefont {Arrigoni}}, \bibinfo {author} {\bibfnamefont {E.}~\bibnamefont {Fradkin}}, \ and\ \bibinfo {author} {\bibfnamefont {S.~A.}\ \bibnamefont {Kivelson}},\ }\href {\doibase 10.1103/PhysRevB.69.214519} {\bibfield  {journal} {\bibinfo  {journal} {Phys. Rev. B}\ }\textbf {\bibinfo {volume} {69}},\ \bibinfo {pages} {214519} (\bibinfo {year} {2004})}\BibitemShut {NoStop}%
\bibitem [{\citenamefont {Mathey}\ and\ \citenamefont {Wang}(2007{\natexlab{a}})}]{Mathey2007polaron}%
  \BibitemOpen
  \bibfield  {author} {\bibinfo {author} {\bibfnamefont {L.}~\bibnamefont {Mathey}}\ and\ \bibinfo {author} {\bibfnamefont {D.-W.}\ \bibnamefont {Wang}},\ }\href {\doibase 10.1103/PhysRevA.75.013612} {\bibfield  {journal} {\bibinfo  {journal} {Phys. Rev. A}\ }\textbf {\bibinfo {volume} {75}},\ \bibinfo {pages} {013612} (\bibinfo {year} {2007}{\natexlab{a}})}\BibitemShut {NoStop}%
\bibitem [{\citenamefont {Lu}\ \emph {et~al.}(2023)\citenamefont {Lu}, \citenamefont {Qu}, \citenamefont {Qi}, \citenamefont {Li},\ and\ \citenamefont {Gong}}]{ShoushuGong2023}%
  \BibitemOpen
  \bibfield  {author} {\bibinfo {author} {\bibfnamefont {X.}~\bibnamefont {Lu}}, \bibinfo {author} {\bibfnamefont {D.-W.}\ \bibnamefont {Qu}}, \bibinfo {author} {\bibfnamefont {Y.}~\bibnamefont {Qi}}, \bibinfo {author} {\bibfnamefont {W.}~\bibnamefont {Li}}, \ and\ \bibinfo {author} {\bibfnamefont {S.-S.}\ \bibnamefont {Gong}},\ }\href {\doibase 10.1103/PhysRevB.107.125114} {\bibfield  {journal} {\bibinfo  {journal} {Phys. Rev. B}\ }\textbf {\bibinfo {volume} {107}},\ \bibinfo {pages} {125114} (\bibinfo {year} {2023})}\BibitemShut {NoStop}%
\bibitem [{\citenamefont {Calabrese}\ and\ \citenamefont {Cardy}(2009)}]{Calabrese_2009}%
  \BibitemOpen
  \bibfield  {author} {\bibinfo {author} {\bibfnamefont {P.}~\bibnamefont {Calabrese}}\ and\ \bibinfo {author} {\bibfnamefont {J.}~\bibnamefont {Cardy}},\ }\href {\doibase 10.1088/1751-8113/42/50/504005} {\bibfield  {journal} {\bibinfo  {journal} {J. Phys. A}\ }\textbf {\bibinfo {volume} {42}},\ \bibinfo {pages} {504005} (\bibinfo {year} {2009})}\BibitemShut {NoStop}%
\bibitem [{\citenamefont {Calabrese}\ and\ \citenamefont {Cardy}(2004)}]{Calabrese_2004}%
  \BibitemOpen
  \bibfield  {author} {\bibinfo {author} {\bibfnamefont {P.}~\bibnamefont {Calabrese}}\ and\ \bibinfo {author} {\bibfnamefont {J.}~\bibnamefont {Cardy}},\ }\href {\doibase 10.1088/1742-5468/2004/06/p06002} {\bibfield  {journal} {\bibinfo  {journal} {J. Stat. Mech.: Theory Exp.}\ }\textbf {\bibinfo {volume} {2004}},\ \bibinfo {pages} {P06002} (\bibinfo {year} {2004})}\BibitemShut {NoStop}%
\bibitem [{\citenamefont {Mattioli}\ \emph {et~al.}(2013)\citenamefont {Mattioli}, \citenamefont {Dalmonte}, \citenamefont {Lechner},\ and\ \citenamefont {Pupillo}}]{spinlessFDD32}%
  \BibitemOpen
  \bibfield  {author} {\bibinfo {author} {\bibfnamefont {M.}~\bibnamefont {Mattioli}}, \bibinfo {author} {\bibfnamefont {M.}~\bibnamefont {Dalmonte}}, \bibinfo {author} {\bibfnamefont {W.}~\bibnamefont {Lechner}}, \ and\ \bibinfo {author} {\bibfnamefont {G.}~\bibnamefont {Pupillo}},\ }\href {\doibase 10.1103/PhysRevLett.111.165302} {\bibfield  {journal} {\bibinfo  {journal} {Phys. Rev. Lett.}\ }\textbf {\bibinfo {volume} {111}},\ \bibinfo {pages} {165302} (\bibinfo {year} {2013})}\BibitemShut {NoStop}%
\bibitem [{\citenamefont {Dalmonte}\ \emph {et~al.}(2015)\citenamefont {Dalmonte}, \citenamefont {Lechner}, \citenamefont {Cai}, \citenamefont {Mattioli}, \citenamefont {L\"auchli},\ and\ \citenamefont {Pupillo}}]{spinlessFDD33}%
  \BibitemOpen
  \bibfield  {author} {\bibinfo {author} {\bibfnamefont {M.}~\bibnamefont {Dalmonte}}, \bibinfo {author} {\bibfnamefont {W.}~\bibnamefont {Lechner}}, \bibinfo {author} {\bibfnamefont {Z.}~\bibnamefont {Cai}}, \bibinfo {author} {\bibfnamefont {M.}~\bibnamefont {Mattioli}}, \bibinfo {author} {\bibfnamefont {A.~M.}\ \bibnamefont {L\"auchli}}, \ and\ \bibinfo {author} {\bibfnamefont {G.}~\bibnamefont {Pupillo}},\ }\href {\doibase 10.1103/PhysRevB.92.045106} {\bibfield  {journal} {\bibinfo  {journal} {Phys. Rev. B}\ }\textbf {\bibinfo {volume} {92}},\ \bibinfo {pages} {045106} (\bibinfo {year} {2015})}\BibitemShut {NoStop}%
\bibitem [{\citenamefont {Kane}\ \emph {et~al.}(2017)\citenamefont {Kane}, \citenamefont {Stern},\ and\ \citenamefont {Halperin}}]{spinlessFDD34}%
  \BibitemOpen
  \bibfield  {author} {\bibinfo {author} {\bibfnamefont {C.~L.}\ \bibnamefont {Kane}}, \bibinfo {author} {\bibfnamefont {A.}~\bibnamefont {Stern}}, \ and\ \bibinfo {author} {\bibfnamefont {B.~I.}\ \bibnamefont {Halperin}},\ }\href {\doibase 10.1103/PhysRevX.7.031009} {\bibfield  {journal} {\bibinfo  {journal} {Phys. Rev. X}\ }\textbf {\bibinfo {volume} {7}},\ \bibinfo {pages} {031009} (\bibinfo {year} {2017})}\BibitemShut {NoStop}%
\bibitem [{\citenamefont {Hohenadler}\ \emph {et~al.}(2012)\citenamefont {Hohenadler}, \citenamefont {Wessel}, \citenamefont {Daghofer},\ and\ \citenamefont {Assaad}}]{spinlessFDD41}%
  \BibitemOpen
  \bibfield  {author} {\bibinfo {author} {\bibfnamefont {M.}~\bibnamefont {Hohenadler}}, \bibinfo {author} {\bibfnamefont {S.}~\bibnamefont {Wessel}}, \bibinfo {author} {\bibfnamefont {M.}~\bibnamefont {Daghofer}}, \ and\ \bibinfo {author} {\bibfnamefont {F.~F.}\ \bibnamefont {Assaad}},\ }\href {\doibase 10.1103/PhysRevB.85.195115} {\bibfield  {journal} {\bibinfo  {journal} {Phys. Rev. B}\ }\textbf {\bibinfo {volume} {85}},\ \bibinfo {pages} {195115} (\bibinfo {year} {2012})}\BibitemShut {NoStop}%
\bibitem [{\citenamefont {Gotta}\ \emph {et~al.}(2021{\natexlab{c}})\citenamefont {Gotta}, \citenamefont {Mazza}, \citenamefont {Simon},\ and\ \citenamefont {Roux}}]{fpairingPhysRevResearch2021}%
  \BibitemOpen
  \bibfield  {author} {\bibinfo {author} {\bibfnamefont {L.}~\bibnamefont {Gotta}}, \bibinfo {author} {\bibfnamefont {L.}~\bibnamefont {Mazza}}, \bibinfo {author} {\bibfnamefont {P.}~\bibnamefont {Simon}}, \ and\ \bibinfo {author} {\bibfnamefont {G.}~\bibnamefont {Roux}},\ }\href {\doibase 10.1103/PhysRevResearch.3.013114} {\bibfield  {journal} {\bibinfo  {journal} {Phys. Rev. Res.}\ }\textbf {\bibinfo {volume} {3}},\ \bibinfo {pages} {013114} (\bibinfo {year} {2021}{\natexlab{c}})}\BibitemShut {NoStop}%
\bibitem [{\citenamefont {Klemmer}\ \emph {et~al.}(2024)\citenamefont {Klemmer}, \citenamefont {Fleper}, \citenamefont {Jonas}, \citenamefont {Sheikhan}, \citenamefont {Kollath}, \citenamefont {K\"ohl},\ and\ \citenamefont {Bergschneider}}]{exp2024pairTunneling}%
  \BibitemOpen
  \bibfield  {author} {\bibinfo {author} {\bibfnamefont {N.}~\bibnamefont {Klemmer}}, \bibinfo {author} {\bibfnamefont {J.}~\bibnamefont {Fleper}}, \bibinfo {author} {\bibfnamefont {V.}~\bibnamefont {Jonas}}, \bibinfo {author} {\bibfnamefont {A.}~\bibnamefont {Sheikhan}}, \bibinfo {author} {\bibfnamefont {C.}~\bibnamefont {Kollath}}, \bibinfo {author} {\bibfnamefont {M.}~\bibnamefont {K\"ohl}}, \ and\ \bibinfo {author} {\bibfnamefont {A.}~\bibnamefont {Bergschneider}},\ }\href {\doibase 10.1103/PhysRevLett.133.253402} {\bibfield  {journal} {\bibinfo  {journal} {Phys. Rev. Lett.}\ }\textbf {\bibinfo {volume} {133}},\ \bibinfo {pages} {253402} (\bibinfo {year} {2024})}\BibitemShut {NoStop}%
\bibitem [{\citenamefont {Eckardt}(2017)}]{Eckardt2017RevMod}%
  \BibitemOpen
  \bibfield  {author} {\bibinfo {author} {\bibfnamefont {A.}~\bibnamefont {Eckardt}},\ }\href {\doibase 10.1103/RevModPhys.89.011004} {\bibfield  {journal} {\bibinfo  {journal} {Rev. Mod. Phys.}\ }\textbf {\bibinfo {volume} {89}},\ \bibinfo {pages} {011004} (\bibinfo {year} {2017})}\BibitemShut {NoStop}%
\bibitem [{\citenamefont {F{\"o}lling}\ \emph {et~al.}(2007)\citenamefont {F{\"o}lling}, \citenamefont {Trotzky}, \citenamefont {Cheinet}, \citenamefont {Feld}, \citenamefont {Saers}, \citenamefont {Widera}, \citenamefont {M{\"u}ller},\ and\ \citenamefont {Bloch}}]{Bloch2007pairTunneling}%
  \BibitemOpen
  \bibfield  {author} {\bibinfo {author} {\bibfnamefont {S.}~\bibnamefont {F{\"o}lling}}, \bibinfo {author} {\bibfnamefont {S.}~\bibnamefont {Trotzky}}, \bibinfo {author} {\bibfnamefont {P.}~\bibnamefont {Cheinet}}, \bibinfo {author} {\bibfnamefont {M.}~\bibnamefont {Feld}}, \bibinfo {author} {\bibfnamefont {R.}~\bibnamefont {Saers}}, \bibinfo {author} {\bibfnamefont {A.}~\bibnamefont {Widera}}, \bibinfo {author} {\bibfnamefont {T.}~\bibnamefont {M{\"u}ller}}, \ and\ \bibinfo {author} {\bibfnamefont {I.}~\bibnamefont {Bloch}},\ }\href {\doibase 10.1038/nature06112} {\bibfield  {journal} {\bibinfo  {journal} {Nature}\ }\textbf {\bibinfo {volume} {448}},\ \bibinfo {pages} {1029} (\bibinfo {year} {2007})}\BibitemShut {NoStop}%
\bibitem [{\citenamefont {Ries}\ \emph {et~al.}(2015)\citenamefont {Ries}, \citenamefont {Wenz}, \citenamefont {Z\"urn}, \citenamefont {Bayha}, \citenamefont {Boettcher}, \citenamefont {Kedar}, \citenamefont {Murthy}, \citenamefont {Neidig}, \citenamefont {Lompe},\ and\ \citenamefont {Jochim}}]{exp2015PairCondensation}%
  \BibitemOpen
  \bibfield  {author} {\bibinfo {author} {\bibfnamefont {M.~G.}\ \bibnamefont {Ries}}, \bibinfo {author} {\bibfnamefont {A.~N.}\ \bibnamefont {Wenz}}, \bibinfo {author} {\bibfnamefont {G.}~\bibnamefont {Z\"urn}}, \bibinfo {author} {\bibfnamefont {L.}~\bibnamefont {Bayha}}, \bibinfo {author} {\bibfnamefont {I.}~\bibnamefont {Boettcher}}, \bibinfo {author} {\bibfnamefont {D.}~\bibnamefont {Kedar}}, \bibinfo {author} {\bibfnamefont {P.~A.}\ \bibnamefont {Murthy}}, \bibinfo {author} {\bibfnamefont {M.}~\bibnamefont {Neidig}}, \bibinfo {author} {\bibfnamefont {T.}~\bibnamefont {Lompe}}, \ and\ \bibinfo {author} {\bibfnamefont {S.}~\bibnamefont {Jochim}},\ }\href {\doibase 10.1103/PhysRevLett.114.230401} {\bibfield  {journal} {\bibinfo  {journal} {Phys. Rev. Lett.}\ }\textbf {\bibinfo {volume} {114}},\ \bibinfo {pages} {230401} (\bibinfo {year} {2015})}\BibitemShut {NoStop}%
\bibitem [{\citenamefont {Kuklov}\ and\ \citenamefont {Moritz}(2007)}]{multimerDetect2007}%
  \BibitemOpen
  \bibfield  {author} {\bibinfo {author} {\bibfnamefont {A.}~\bibnamefont {Kuklov}}\ and\ \bibinfo {author} {\bibfnamefont {H.}~\bibnamefont {Moritz}},\ }\href {\doibase 10.1103/PhysRevA.75.013616} {\bibfield  {journal} {\bibinfo  {journal} {Phys. Rev. A}\ }\textbf {\bibinfo {volume} {75}},\ \bibinfo {pages} {013616} (\bibinfo {year} {2007})}\BibitemShut {NoStop}%
\bibitem [{\citenamefont {Carusotto}\ and\ \citenamefont {Castin}(2005)}]{corrSCexp2005}%
  \BibitemOpen
  \bibfield  {author} {\bibinfo {author} {\bibfnamefont {I.}~\bibnamefont {Carusotto}}\ and\ \bibinfo {author} {\bibfnamefont {Y.}~\bibnamefont {Castin}},\ }\href {\doibase 10.1103/PhysRevLett.94.223202} {\bibfield  {journal} {\bibinfo  {journal} {Phys. Rev. Lett.}\ }\textbf {\bibinfo {volume} {94}},\ \bibinfo {pages} {223202} (\bibinfo {year} {2005})}\BibitemShut {NoStop}%
\bibitem [{\citenamefont {Song}\ \emph {et~al.}(2024)\citenamefont {Song}, \citenamefont {Lou},\ and\ \citenamefont {Chen}}]{QiSong2024CSF}%
  \BibitemOpen
  \bibfield  {author} {\bibinfo {author} {\bibfnamefont {Q.}~\bibnamefont {Song}}, \bibinfo {author} {\bibfnamefont {J.}~\bibnamefont {Lou}}, \ and\ \bibinfo {author} {\bibfnamefont {Y.}~\bibnamefont {Chen}},\ }\href {\doibase 10.1103/PhysRevA.110.063312} {\bibfield  {journal} {\bibinfo  {journal} {Phys. Rev. A}\ }\textbf {\bibinfo {volume} {110}},\ \bibinfo {pages} {063312} (\bibinfo {year} {2024})}\BibitemShut {NoStop}%
\bibitem [{\citenamefont {Roux}\ \emph {et~al.}(2009)\citenamefont {Roux}, \citenamefont {Capponi}, \citenamefont {Lecheminant},\ and\ \citenamefont {Azaria}}]{SvN2009spin}%
  \BibitemOpen
  \bibfield  {author} {\bibinfo {author} {\bibfnamefont {G.}~\bibnamefont {Roux}}, \bibinfo {author} {\bibfnamefont {S.}~\bibnamefont {Capponi}}, \bibinfo {author} {\bibfnamefont {P.}~\bibnamefont {Lecheminant}}, \ and\ \bibinfo {author} {\bibfnamefont {P.}~\bibnamefont {Azaria}},\ }\href {\doibase https://doi.org/10.1140/epjb/e2008-00374-7} {\bibfield  {journal} {\bibinfo  {journal} {The European Physical Journal B}\ }\textbf {\bibinfo {volume} {68}},\ \bibinfo {pages} {293} (\bibinfo {year} {2009})}\BibitemShut {NoStop}%
\bibitem [{\citenamefont {Moreno}\ \emph {et~al.}(2011)\citenamefont {Moreno}, \citenamefont {Muramatsu},\ and\ \citenamefont {Manmana}}]{tJMoreno2011}%
  \BibitemOpen
  \bibfield  {author} {\bibinfo {author} {\bibfnamefont {A.}~\bibnamefont {Moreno}}, \bibinfo {author} {\bibfnamefont {A.}~\bibnamefont {Muramatsu}}, \ and\ \bibinfo {author} {\bibfnamefont {S.~R.}\ \bibnamefont {Manmana}},\ }\href {\doibase 10.1103/PhysRevB.83.205113} {\bibfield  {journal} {\bibinfo  {journal} {Phys. Rev. B}\ }\textbf {\bibinfo {volume} {83}},\ \bibinfo {pages} {205113} (\bibinfo {year} {2011})}\BibitemShut {NoStop}%
\bibitem [{\citenamefont {Danshita}\ and\ \citenamefont {Mathey}(2013)}]{PhysRevA.87.021603}%
  \BibitemOpen
  \bibfield  {author} {\bibinfo {author} {\bibfnamefont {I.}~\bibnamefont {Danshita}}\ and\ \bibinfo {author} {\bibfnamefont {L.}~\bibnamefont {Mathey}},\ }\href {\doibase 10.1103/PhysRevA.87.021603} {\bibfield  {journal} {\bibinfo  {journal} {Phys. Rev. A}\ }\textbf {\bibinfo {volume} {87}},\ \bibinfo {pages} {021603(R)} (\bibinfo {year} {2013})}\BibitemShut {NoStop}%
\bibitem [{\citenamefont {Mathey}\ and\ \citenamefont {Wang}(2007{\natexlab{b}})}]{PhysRevA.75.013612}%
  \BibitemOpen
  \bibfield  {author} {\bibinfo {author} {\bibfnamefont {L.}~\bibnamefont {Mathey}}\ and\ \bibinfo {author} {\bibfnamefont {D.-W.}\ \bibnamefont {Wang}},\ }\href {\doibase 10.1103/PhysRevA.75.013612} {\bibfield  {journal} {\bibinfo  {journal} {Phys. Rev. A}\ }\textbf {\bibinfo {volume} {75}},\ \bibinfo {pages} {013612} (\bibinfo {year} {2007}{\natexlab{b}})}\BibitemShut {NoStop}%
\end{thebibliography}%
\end{document}